\documentclass[prc,showpacs,showkeys,onecolumn,preprint]{revtex4-1}

\usepackage{amssymb}
\usepackage{amsmath}
\usepackage{epsfig}
\usepackage{epstopdf}
\usepackage{color}
\usepackage{slashed}
\newcommand{\be}{\begin{equation}}
\newcommand{\ee}{\end{equation}}
\newcommand{\ben}{\begin{eqnarray}}
\newcommand{\een}{\end{eqnarray}}
\newcommand{\lb}{\label}

\usepackage{longtable}
\usepackage{booktabs}

\usepackage{graphicx}
\usepackage{hyperref}
\usepackage{xcolor}
\hypersetup{colorlinks=true, linkcolor=blue, citecolor=blue}

\begin{document}
\title{Finite-volume and magnetic effects on the phase structure of the three-flavor Nambu--Jona-Lasinio model}
%%%%%%%%%%%%%%%%%%%%%%%%%%%%%%%%%%%%%
%%%%%%%%%%%%%%%%%%%%%%%%%%%%%%%%%%%%%%%%%%
\author{Luciano M. Abreu}
\email[]{luciano.abreu@ufba.br}
\affiliation{Instituto de F\'{\i}sica, Universidade Federal da Bahia, 40170-115, Salvador, BA, Brazil}
\author{Emerson B. S. Corr\^ea}
\email[]{emersoncbpf@gmail.com}
\affiliation{Faculdade de F\'isica, Universidade Federal do Sul e Sudeste do Par\'a, 68505-080, Marab\'a, PA, Brazil}
\author{Cesar A. Linhares}
\email[]{linharescesar@gmail.com}
\affiliation{Instituto de F\'{\i}sica, Universidade do Estado do Rio de Janeiro, 20559-900, Rio de Janeiro, RJ, Brazil}
\author{Adolfo P. C. Malbouisson}
\email[]{adolfo@cbpf.br}
\affiliation{Centro Brasileiro de Pesquisas F\'{\i}sicas/MCTI, 22290-180, Rio de Janeiro, RJ, Brazil}
%
%\begin{center}
%	{\textit{Dedicated to the memory of Edson B. S. Corr\^{e}a}}
%\end{center}
%%%%%%%%%%%%%%%%%%%%%%%%%%%%%%%%%%%%%%%%%%%%%%%%%%%%%%%%%%%%%
%%%%%%%%%%%%%%%%%%%%%%%%%%%%%%%%%%%%%%%%%%%%%%%%%%%%%%%%%%%%%
\begin{abstract}

In this work we analyze the finite-volume and magnetic effects on the phase structure 
of a generalized version of Nambu--Jona-Lasinio model with three quark flavors. By making use of mean-field approximation and Schwinger's proper-time method in a toroidal topology with antiperiodic conditions, we investigate the gap equation solutions under the change of the size of compactified coordinates, strength of magnetic field, temperature and chemical potential. The 't Hooft interaction contributions are also evaluated. 
The thermodynamic behavior is strongly affected by the combined effects of relevant variables. The findings suggest that  the broken phase is disfavored due to both increasing of temperature and chemical potential, and the drop of the cubic volume of size $L$, whereas it is stimulated with the augmentation of magnetic field. In particular, the reduction of $L$ (remarkably at $L\approx 0.5 - 3 $~fm) engenders a reduction of the constituent masses for $u,d,s$-quarks through a crossover phase transition to the their corresponding current quark masses. 
On the other hand, the presence of a magnetic background generates greater values constituent quark masses, inducing smaller sizes and greater temperatures at which the constituent quark masses drop to the respective current ones.  

\end{abstract}

\maketitle
%
%%%%%%%%%%%%%%%%%%%%%%%%%%%%%%%%%%%%%%%%%%%%%%%%%%%%%%%%%%%%%
%%%%%%%%%%%%%%%%%%%%%%%%%%%%%%%%%%%%%%%%%%%%%%%%%%%%%%%%%%%%%
%%%%%%%%%%%%%%%%%%%%%%%%%%%%%%%%%%%%%%%%%%%%%%%%%%%%%%%%%%%%%
%%%%%%%%%%%%%%%%%%%%%%%%%%%%%%%%%%%%%%%%%%%%%%%%%%%%%%%%%%%%%

\section{Introduction}

%%%%%%%%%%%%%%%%%%%%%%%%%%%%%%%%%%%%%%%%%%%%%%%%%%%%%%%%%%%%%
%%%%%%%%%%%%%%%%%%%%%%%%%%%%%%%%%%%%%%%%%%%%%%%%%%%%%%%%%%%%%
%%%%%%%%%%%%%%%%%%%%%%%%%%%%%%%%%%%%%%%%%%%%%%%%%%%%%%%%%%%%%
%%%%%%%%%%%%%%%%%%%%%%%%%%%%%%%%%%%%%%%%%%%%%%%%%%%%%%%%%%%%%

In the recent decades we have witnessed experimental and theoretical progresses in the assessment of strongly interacting matter under extreme conditions. One of the most interesting predictions of its underlying theory, Quantum Chromodynamics (QCD), is that it experiences a phase transition to a deconfined state at sufficiently 
high temperatures. This mentioned state composed of quarks and gluons is commonly referred to as the quark-gluon plasma (QGP) and it 
can now be observed in heavy-ion collisions~\cite{qgpdisc,rev-qgp}. Its thermodynamic and transport properties continues to the forefront of endeavour of community  to outline its rich phase structure~\cite{rev-qgp,Prino:2016cni,Pasechnik:2016wkt}.

In particular, one interesting feature on the phase structure of strongly interacting matter is the finite-volume effects. There are estimations indicating that QGP-like systems yielded in heavy-ion collisions are of the order of units or dozens of fermi~\cite{Bass:1998qm,Palhares:2009tf,Graef:2012sh}. In this sense, the influence of the size of the system on its thermodynamic behaviour and phase diagram has been widely analyzed in the literature through distinct and effective approaches, as Dyson--Schwinger equations of QCD~\cite{Luecker:2009bs,Li:2017zny,Shi:2018swj}, quark-meson model~\cite{Braun:2004yk,Braun:2005fj},  extensions and generalizations of the Nambu--Jona-Lasinio (NJL) model~\cite{Ferrer:1999gs,Abreu:2006,Ebert0,Abreu:2009zz,Abreu:2011rj,Bhattacharyya:2012rp,Bhattacharyya:2014uxa,Bhattacharyya2,Pan:2016ecs,Kohyama:2016fif,Wang:2018qyq}, and others~\cite{Gasser:1986vb,Damgaard:2008zs,Fraga,Abreu3,Abreu6,Ebert3,Abreu4,Abreu5,Abreu7,Bao1,PhysRevC.96.055204,Samanta,Wu,Klein:2017shl,Shi}. The main findings of these works point out that the finite volume, combined  with other relevant variables, act on the thermodynamic behavior of strongly interacting matter.

In the papers cited above, important questions arise in order to have a detailed description of limited-size physical systems, like the mentioned QGP. One example is the spontaneous symmetry breaking of chiral symmetry, which occurs for systems in the bulk approximation, but is disfavored for finite volume. Thus, this context gives rise to the natural debate about what are the conditions in which an ideal bulk system seems a good approximation for confined systems. In other words, one may wonder about the range of the size of the system $V=L^3$ in which the finite-volume effects affect the phase structure and the spontaneous symmetry breaking, making modifications in the thermodynamical properties of quarkionic matter. 

Specially, in a very recent paper by Wang et al.~\cite{Wang:2018qyq} is discussed the influence of the finite-volume effects on the chiral phase transition of quark matter at finite temperature, based on a two-flavour NJL model approach with a proper-time regularization and use of a stationary-wave condition.  It is found that when the cubic volume size $L$ is larger than 500 fm, the chiral quark condensate is indistinguishable from that at $L = \infty$, which is far greater than the estimates of the size of QGP produced at laboratory and the lattice QCD simulations. Also, they found that when the space size $L$ is less than 0.25 fm, the spontaneous symmetry breaking concept is no longer valid.

On the other hand, the importance of a magnetic background on the phase diagram of strongly interacting matter has also been a subject of great interest. The motivation comes from both phenomenology of heavy ion collisions and compact stars, in which a strong magnetic background is produced~\cite{Kharzeev,Skokov:2009qp,Chernodub:2010qx,Ayala1,Tobias,Heber,MAO,Ayala2,Mamo:2015dea,Pagura,Magdy,Ayala0}. Taking into account the case of heavy-ion collisions at RHIC and LHC, a notable dependence of phase diagram with the strength of magnetic field $e H$ is estimated in the hadronic scale, i.e. $e H \sim 1-10\; m_{\pi}^2$ ($ m_{\pi} = 140$ MeV is the pion mass). In this scenario, intriguing distinct phenomena appearing for different ranges of temperatures have been proposed recently by use of effective approaches: for sufficiently small magnetic field, magnetic catalysis (enhancement of broken phase) occurs at smaller temperatures, while the inverse magnetic catalysis effect (stimulation of the restoration of chiral symmetry) takes place at higher temperatures~\cite{Tobias,MAO,Mamo:2015dea,Pagura,Magdy,Ayala0}. It should be also noted that inverse magnetic catalysis can also occur even at $T=0$, in the region of the phase diagram involving chemical potential versus $e H$ for intermediate/high values of field strength.

Consequently, a natural discussion emerges about the combined effects on the phase structure of a thermal gas of quark matter (which might be described as a first approximation by the NJL model) restricted to a reservoir and under the effect of a magnetic background.

Hence, in view of these recent theoretical and experimental advances    
on the physics of strongly interacting matter, we believe that there is still room to other contributions. To this end, the main interest of this paper is to extend the analysis performed in Ref.~\cite{Wang:2018qyq}. In the present work we will address the questions raised above under the framework of a generalized version of NJL model. We investigate the finite-size effects on the phase structure of three-flavor NJL model with 't Hooft interaction, which engenders a six-fermion interaction term, without and with the presence of a magnetic background. We make use of mean-field approximation and Schwinger's proper-time method in a toroidal topology with antiperiodic conditions, manifested by the utilization of generalized Matsubara prescription for the imaginary time and spatial coordinate compactifications. To deal with the nonrenormalizable nature of the NJL model, the ultraviolet cut-off regularization procedure is employed. The behavior of constituent quark masses $M_{u}, M_d $ and $ M_s$, engendered by the solutions of gap equations, are studied under the change of the size $ L$ of compactified coordinates, temperature $T$ and chemical potential $\mu$. The 't Hooft interaction contributions are also evaluated. 

We organize the paper as follows. In Section~II, we calculate the $(T,L,\mu,H)$-dependent
effective potential and gap equations  obtained from the three-flavor NJL model in the mean-field
approximation, using Schwinger's proper-time method generalized Matsubara prescription. The phase structure of the system and the behavior of constituent quark masses $M_{u}, M_d $ and $M_s$ are shown and analyzed in Section~III, without and with the presence of a magnetic background. Finally, Section~IV presents some concluding remarks.

%%%%%%%%%%%%%%%%%%%%%%%%%%%%%%%%%%%%%%%%%%%%%%%%%%%%%%%%%%%%%%%%%%%%%%%%%%%%%%%%%%%%%%%%%%%%%%%%%%%%%%%%%%%%%%%%%%%%%%%%%%%%%%%%%%%%%%%%%%%%%%%%%%%%%%%%%%%%%%%%%%%%%%%%
\section{The Model}
%%%%%%%%%%%%%%%%%%%%%%%%%%%%%%%%%%%%%%%%%%%%%%%%%%%%%%%%%%%%%%%%%%%%%%%%%%%%%%%%%%%%%%%%%%%%%%%%%%%%%%%%%%%%%%%%%%%
%%%%%%%%%%%%%%%%%%%%%%%%%%%%%%%%%%%%%
%%%%%%%%%%%%%%%%%%%%%%%%%%%%%%%%%%%%%
\subsection{Lagrangian and thermodynamics}
%%%%%%%%%%%%%%%%%%%%%%%%%%%%%%%%%%%%%
%%%%%%%%%%%%%%%%%%%%%%%%%%%%%%%%%%%%%
	
%\begin{center}
	%{\bf{{II - The Model}}}
%\end{center}
Let us start by introducing the $N_f$-flavor version of the NJL model, whose most commonly used Lagrangian density reads~\cite{NJL,NJL1,Vogl,Klevansky,Hatsuda,Buballa},  
\begin{eqnarray}
\mathcal{L}_{\rm NJL} = \bar{q}\,(i{\slashed{\partial}})\,q + \mathcal{L}_{\rm Mass} +  \mathcal{L}_{4} +  \mathcal{L}_{6},
\label{L}
\end{eqnarray}
where $q $ is the quark field carrying $N_f $ flavors; $\mathcal{L}_{\rm Mass} $ denotes the mass term, 
\begin{eqnarray}
\mathcal{L}_{\rm Mass} = -  \bar{q}\, \hat{m} \,q ,
\label{Lmass}
\end{eqnarray}
with $\hat{m} = {\rm diag}(m_{1}, \ldots ,m_{N_f})$ being the corresponding
mass matrix; $\mathcal{L}_{4} $ represents the four-fermion-interaction
term, 
\begin{eqnarray}
\mathcal{L}_{4} = G\, \sum_{a=0}^{N_f ^2 - 1} \left[\left(\bar{q}\lambda_{a}q\right)^{2}+\left(\bar{q} i\gamma_{5}\lambda_{a}q\right)^{2}\right],
\label{L4}
\end{eqnarray}
with $G$ being the respective coupling constant and $\lambda_a$ the generators of $U(N_f)$ in flavor space; and  $\mathcal{L}_{6} $ designates the so-called 't Hooft interaction, 
\begin{eqnarray}
\mathcal{L}_{6} = -\kappa \left[\det\left(\bar{q}(1-\gamma_{5})q\right)+\det\left(\bar{q}(1+\gamma_{5})q\right)\right],
\label{L6}
\end{eqnarray}
with $\kappa$ being the corresponding coupling constant.

We remark some features of the model introduced above. The version of NJL model presented here is generalized to any number of $N_f$ flavors, but we are particularly interested in the case of $N_f = 3$. This means that the quark field has its components related to the lightest quark flavors: $q = (u, d, s)^T $.  This allows to identify the generators  $\lambda_a$  of $U(3)$ in flavor space as follows: $(\lambda _1, \dots , \lambda _8)$ are the Gell-Mann matrices and $\lambda_{0} = \sqrt{2/3}I_{3}$. Also, the corresponding mass matrix becomes $\hat{m}  \equiv diag(m_{u}, m_d ,m_{s})$. Besides, henceforth the isospin symmetry on the Lagrangian level is assumed, i.e. 
$m_u = m_d$,  while $SU(3)$-flavor symmetry is explicitly broken, generating $m_s \neq m_u$. 

From the perspective of $U(N_f)_L \times U(N_f)_R$-symmetry in Lagrangian terms, we see that $\mathcal{L}_{4}$ is a symmetric four-point
interaction. On the other hand, the 't Hooft interaction $\mathcal{L}_{6}$ is a determinant in flavor space, which means that it is a flavor-mixing $2N_f $-point interaction, involving an incoming and an outgoing quark of each flavor. Thus, for three flavors it engenders a six-fermion-interaction term, and is $SU(N_f )_L \times SU(N_f )_R $-symmetric, but breaks the $U_A(1)$ symmetry which was left unbroken by $\mathcal{L}_{4}$. Thus, $\mathcal{L}_{6}$ express the $U_A(1)$ anomaly, which appears in the QCD context from the gluon sector, in terms of a tree-level interaction in the present pure quark model. As discussed in Refs.~\cite{Klevansky,Hatsuda,Buballa}, the 't Hooft interaction is phenomenologically relevant to yield the mass splitting of $\eta $ and $\eta ^{\prime}$ mesons. Moreover, it can be shown that in the chiral limit,  $m_u = m_d =m_s = 0$, the $\eta ^{\prime}$ acquires a finite value of mass due to  $\mathcal{L}_{6}$, whereas the other pseudoscalar mesons stay massless.

The focus of the present analysis concerns the lowest-order estimate of the phase structure of this model. In this sense, the calculations for obtaining relevant quantities will be performed within the mean-field (Hartree) approximation. To this end, the non-vanishing quark condensates,
\be
\phi_{i} \equiv \left\langle  \bar{q}_i q_i \right\rangle 
\label{phi1-ini}
\ee
($i= u,d,s$), are assumed to be the only allowed expectation values which are bilinear in the quark fields. In this approximation, the interaction terms in $L_{\rm NJL}$ are linearized in the presence of $\phi _i$, that is,
 \be 
 (\bar{q} q) ^2 \simeq 2 \phi _i (\bar{q}_i q_i) - \phi _i ^2,
\label{phi2-ini}
\ee 
which means that terms quadratic in the fluctuations will be neglected. Moreover, terms in channels without condensate or nondiagonal in flavor space, like $\left(\bar{q}\lambda_{b} q\right) $, $b \neq 0,$ and $\left(\bar{q} i\gamma_{5}\lambda_{a}q\right)$, are are excluded. 
Therefore, in this context the NJL Lagrangian density can be rewritten as 
\begin{eqnarray}
\mathcal{L}_{\rm MF}=\bar{q}\left( i{\slashed{\partial}} - M \right)q - 2G (\phi_{u}^{2}+\phi_{d}^{2}+\phi_{s}^{2})+4\kappa \phi_{u}\phi_{d}\phi_{s},
\label{NJLH}
\end{eqnarray}
where we have introduced $M $ as the constituent quark mass matrix, which is diagonal in flavor space: $ M \equiv {\rm diag}(M_{u}, M_d , M_{s})$. Notice that the constant terms in $\mathcal{L}_{\rm MF}$ have been neglected, since they give trivial contributions.

At this point we can present the thermodynamic potential density at temperature $T$ and quark chemical potential $\mu$, which is defined by
\ben
\Omega (T, \mu) &  = & 
- \frac{  T  }{V}  \ln{\mathcal{Z}} \nonumber \\
& = & - \frac{  1 }{\beta V } \ln{ {\rm Tr} \exp{ \left[ -\beta \int d^3 x \left( \mathcal{H} - \mu q^{\dagger } q \right)\right] } } , 
\lb{effpot1} 
\een 
where $\mathcal{Z}$ is the grand canonical partition function, $\beta = 1/ T $, $\mathcal{H}$ the Hamiltonian density (the Euclidean version of Lagrangian density $\mathcal{L}_{\rm MF}$) and ${\rm Tr}$ the functional trace over all states of the system (spin, flavor, color and momentum). Thus, the integration over fermion fields generates the mean-field thermodynamic potential in the form
\ben
\Omega (T, \mu) &  = &  2G (\phi_{u}^{2}+\phi_{d}^{2}+\phi_{s}^{2})- 4\kappa \phi_{u}\phi_{d}\phi_{s} \nonumber \\
& &  + \sum_ {i=u,d,s}\Omega _{M_i} \left( T, \mu_i \right), 
\lb{effpot2} 
\een 
where $\Omega _{M_i} \left( T, \mu_i \right)$ is the free Fermi-gas contribution, 
\ben
\Omega  _{M_i} \left( T, \mu_i \right) = - \frac{1}{\beta} \sum _{n_{\tau}} \int \frac{d^3p}{(2\pi)^3} {\rm tr} \ln{\left[ \slashed{p} \; 1_i - \mu _i \gamma ^0 - M_i \right]}. 
\label{effpot3} 
\een 
Here the sum over $n_{\tau}$ denotes the sum over the fermionic Matsubara frequencies, $p ^0 = i \omega _{n_{\tau} } = \left( 2 n_{\tau} + 1 \right) \pi / \beta ; \; (n_{\tau} = 0, \pm 1, \pm 2 , \ldots)$.

Then, the minimization of the thermodynamic potential in Eq.~(\ref{effpot3}) with respect to quark condensates allows us to obtain the following gap equations,
\begin{eqnarray}
M_{i} = m_{i} - 4G \phi_{i} + 2\kappa \phi_{j} \phi_{k} \;\;(i\neq j \neq k). 
\label{massa}
\end{eqnarray}
The physical solutions of Eq.~(\ref{massa}) are determined from the stationary points of the thermodynamic potential, which lead to the standard expression for the quark condensates, 
\begin{eqnarray}
\phi_{i} \equiv \left\langle  \bar{q}_i q_i \right\rangle  = - 4 N_{c} M_i \frac{1}{\beta} \sum _{n_{\tau}}  \int \frac{d^3 p}{(2 \pi)^3}\frac{1}{\tilde{\omega}_{n_{\tau}} ^{2} + \vec{p}^{2} + M^{2}_{i}},
\label{condensado}
\end{eqnarray}
where $N_c = 3$ and  
\be
 \tilde{\omega}_{n_{\tau}} = \frac{2\pi}{\beta }
	\left( n_{\tau} + \frac{1}{2} - i\frac{\mu \beta}{2\pi}\right). 
\ee 
We have assumed for simplicity the quark chemical potentials as $\mu _i \equiv \mu $.
Eq.~(\ref{massa}) contains a non-flavor mixing term proportional to the
coupling constant $G$ (coming from the four-point interaction contribution) and a flavor-mixing term proportional to the coupling constant $\kappa$ (coming from the six-point interaction).
Hence, the role of the 't Hooft term is to engender flavor-mixing contributions in the constituent masses.

%%%%%%%%%%%%%%%%%%%%%%%%%%%%%%%%%%%%%
%%%%%%%%%%%%%%%%%%%%%%%%%%%%%%%%%%%%%
\subsection{Generalized Matsubara prescription and proper-time formalism}
%%%%%%%%%%%%%%%%%%%%%%%%%%%%%%%%%%%%%
%%%%%%%%%%%%%%%%%%%%%%%%%%%%%%%%%%%%%
\label{Gen-Matsubara}

To take into account finite-size effects on the phase structure of the model, we denote the Euclidean coordinate vectors by $x_E=(x_{\tau},x_1,x_2,x_3)$, where $x_{\tau}\in[0,\beta]$ and $x_j\in[0,L_j] \; (j=1,2,3)$ , with $L_j$ being the length of the compactified spatial dimensions.  As a consequence, the Feynman rules explicited in the arguments of the sum-integral mixing in Eq.~(\ref{condensado}) must be replaced according to the generalized Matsubara prescription~\cite{livro,PR2014,Emerson}, i.e.,
\begin{eqnarray}
\frac{1}{\beta }\sum_{ n_{\tau}=-\infty}^{\infty} \int\frac{d^3p}{(2\pi)^3}f(\tilde{\omega}_{n_{\tau}},\vec{p})\rightarrow \frac{1}{\beta L_1 L_2 L_3}\sum_{ n_{\tau} ,n_1,n_2,n_3=-\infty}^{\infty} f \left( \tilde{\omega}_{n_{\tau}}, \bar{\omega}_{n_1}, \bar{\omega}_{n_2},\bar{\omega}_{n_3} \right),\label{feynmanrule}
\end{eqnarray}
such that
\begin{eqnarray}
 {p}_{j}\rightarrow \bar{\omega} _{n_j} \equiv \frac{2\pi}{L_{j}}
	\left(n_{j} + b_{j}\right) \,,  \label{Matsubara}
\end{eqnarray}
where $n_{\tau}, n_{\alpha} = 0,\pm 1 , \pm 2, \ldots$ Due to the KMS conditions~\cite{livro,PR2014,Bellac,Kapusta}, which determines the thermal Matsubara frequencies according to the field statistics, the fermionic nature of the system under study imposes an antiperiodic condition in the imaginary-time coordinate.
 
However, in the case of spatial compactified coordinates $x_j$ there are no restrictions with respect to the periodicity. So, the choice of the boundary conditions of the fermion fields in the spatial directions results in a relevant issue, especially because it affects the meson masses and other relevant observables, as claimed in studies mentioned in Introduction. Current lattice QCD simulations usually employ periodic quark boundary conditions~\cite{Klein:2017shl},  thanks to its tendency to reduce the finite volume effects, favoring the obtention of the thermodynamic limit.   
Nevertheless, as has been supported by a large number of studies that make use of effective QCD models~\cite{Gasser:1986vb,Klein:2017shl,Shi:2018swj,Wang:2018qyq}, the fields should take the same boundary condition in their spatial and temporal directions. 
One important aftermath of this selection is the symmetry in the formalism between the spatial and temporal directions, through the exchange $L_i \leftrightarrow \beta $, causing $T,L_i$-independent coupling constants $G$ and $\kappa$. So, we can fit the model parameters at zero temperature and infinite volume and hold them in the  finite temperature and volume analysis.   
Thus, following this assumption, we also adopt antiperiodic boundary conditions for the compactified spatial coordinates: $b_{j}$ in Eq.~(\ref{Matsubara}) assumes the value $1/2$.

%%%%%%%%%%5%
In the present work the thermodynamic potential and the gap equations will be treated within the Schwinger proper-time method~\cite{Schwinger,Farina}.  Accordingly, the kernel of the propagator in Eq.~({\ref{condensado}}) can be rewritten as 
\begin{eqnarray}
	\frac{1}{\tilde{\omega}_{n_{\tau}} ^{2}  +\vec{p}^{\,2}+M_{i}^{2}}=	\int_{0}^{\infty} d\tau \exp\left[-\tau \left( \tilde{\omega}_{n_{\tau}} ^{2} +\vec{p}^{\,2}+M_{i}^{2} \right) \right],
	\label{proptime}
\end{eqnarray}
where $\tau$ is the so-called proper time. Therefore, the application of  Eq.~({\ref{proptime}}) and the generalized Matsubara prescription~(\ref{Matsubara}) into ({\ref{condensado}}) yields the quark chiral condensates expressed as
\begin{eqnarray}
\phi_{i} &=& -\frac{4N_{c} M_{i}}{\beta L_{x}L_{y}L_{z}}\int_{0}^{\infty} d\tau \exp[-\tau(M_{i}^{2}-\mu^{2})]~ \sum_{n_\tau,n_x,n_y,n_z = -\infty}^{+\infty} \nonumber \\
&&\times \exp\left[-\tau(4\pi^2/\beta^2)(n_\tau + 1/2)^2 \right] \exp[i(2n_\tau +1)2\pi\mu\tau/\beta] \nonumber \\
&&\times\exp\left[-\tau(4\pi^2/L_{x}^2)(n_x + 1/2)^2 \right]\exp\left[-\tau(4\pi^2/L_{y}^2)(n_y + 1/2)^2 \right] \nonumber \\
&&\times\exp\left[-\tau(4\pi^2/L_{z}^2)(n_z + 1/2)^2 \right].
\label{phi01}
\end{eqnarray}

In addition, to perform the necessary manipulations in a relatively simple and more tractable way, it is convenient to employ the Jacobi theta functions~\cite{Bellman}.  Noticing the very useful property of the second Jacobi theta function,  $\theta_{2} \left[b\,;\,q\right]$,
\begin{eqnarray}
\sum_{n=-\infty}^{+\infty}\exp\left[-a\left(n+1/2\right)^2\right]\exp\left[i\left(2n+1\right)b\right]  &=&2\left[\exp(-a)\right]^{1/4}\sum_{n=0}^{+\infty}\left[\exp(-a)\right]^{n(n+1)}\cos\left[\left(2n+1\right)b\right] \nonumber \\
& =& \theta_{2} \left[b\,;\,\exp(-a)\right], 
\label{theta2}
\end{eqnarray}
then the $(T,L_{j},\mu) $-dependent chiral quark condensate in Eq.~(\ref{phi01}) can be expressed in the following way:
\begin{eqnarray}
\phi_{i}(T,L_{j},\mu) &=& -\frac{4N_{c} M_{i}}{\beta L_{1}L_{2}L_{3}}\int_{0}^{\infty} d\tau \exp[-\tau(M_{i}^{2}-\mu^{2})]\,\theta_{2}\left[\frac{2\pi\mu\tau}{\beta}\,;\,\exp(-4\pi^2 \tau/\beta^2)\right] \nonumber \\
&&\times \theta_{2}\left[0\,;\,\exp(-4\pi^2 \tau/L_{1}^2)\right]\theta_{2}\left[0\,;\,\exp(-4\pi^2 \tau/L_{2}^2)\right]\theta_{2}\left[0\,;\,\exp(-4\pi^2 \tau/L_{3}^2)\right].
\label{phi2}
\end{eqnarray}

It is worthy noticing that in order to regularize the proper-time approach, we introduce a ultraviolet cutoff $\Lambda$, namely,
\begin{eqnarray}
	\int_{0}^{\infty}f(\tau)d\tau \rightarrow \int_{1/\Lambda^2}^{\infty}f(\tau)d\tau.
	\label{cuttof}	
\end{eqnarray}

%%%%%%%%%%%%%%%%%%%%%%%%%%%%%%%%%%%%%%%%%%%%%%%%%%%%%

We conclude this section by remarking that the bulk form of the system can be studied straightforwardly. To do this, the continuum limit is performed by  reverting the Matsubara prescription in Eq.~(\ref{feynmanrule}), which yields in Eq.~(\ref{phi2}) Gaussian integrals in  momentum space. So, the  expression for the chiral quark condensate acquires the form 
\begin{eqnarray}
\phi_{i}(T,L_{j} \rightarrow \infty ,\mu)  & \rightarrow & -\frac{N_{c} M_{i}}{2\sqrt{\pi^{3}}~\beta}\int_{1/\Lambda^2}^{\infty} \frac{d \tau}{\tau^{3/2}} \exp[-\tau(M_{i}^{2}-\mu^{2})]\,\theta_{2}\left[\frac{2\pi\mu\tau}{\beta}\,;\,\exp(-4\pi^2 \tau/\beta^2)\right]. \nonumber \\
\label{phibulk}
\end{eqnarray}

Hence, we have obtained above the thermodynamic potential and gap equations whose solutions produce the $(T,L_{j},\mu) $-dependent constituent quark masses. Other thermodynamic quantities such as pressure, entropy and others can be obtained by similar procedures to those described above.

In the next section, we will discuss the thermodynamic behavior of the present model.

%%%%%%%%%%%%%%%%%%%%%%%%%%%%%%%%%%%%%
%%%%%%%%%%%%%%%%%%%%%%%%%%%%%%%%%%%%%
\subsection{Presence of a magnetic background}
%%%%%%%%%%%%%%%%%%%%%%%%%%%%%%%%%%%%%
%%%%%%%%%%%%%%%%%%%%%%%%%%%%%%%%%%%%%
\label{magnetic-field}

Considering now the system in the presence of an uniform external magnetic field, we redefine the Lagrangian density in Eq.~(\ref{L}) with the derivative in kinetic term replaced by the modified covariant derivative: $\tilde{D}_{\mu} = \partial_{\mu} + i \hat{Q} A^{ext}_{\mu}$, where $A^{ext}_{\mu}$ denotes the four-potential associated to the magnetic background,  and $\hat{Q}$ is the quark charge electric matrix, $ \hat{Q} = \mathrm{diag}[\left( Q_u, Q_d, Q_s \right) e ]$, with $Q_u = - 2 Q_d = -2 Q_s = 2 /3$. We choose for convenience the Landau gauge, $A^{\mu}_{ext}=(0,-Hx_2,0,0)$, where $H$ is the intensity of the external magnetic field in the direction $z$. Therefore, the field operators are given by the set of normalized eigenfunctions of the Landau basis, which means that the energy eingenvalues associated to the solutions of equation of motion read,
\begin{eqnarray}
E_i ^2=p_z^2+M_{i}^2+  |Q_i | \omega \left(2l+1-s\right),\label{eqII9}
\end{eqnarray}
where $\omega \equiv e H $ is the so-called cyclotron frequency; $s=\pm1$ is the spin variable; and $l=0,1,2,...$, denotes the so-called Landau levels.

A fundamental consequence of presence of magnetic background is the modification of  Feynman rules: the four-momentum integrals suffer the dimensional reduction, i.e.
\begin{eqnarray}
\int\frac{d^4p}{(2\pi)^4}f(p)\rightarrow\frac{ |Q_i | \omega }{2\pi} \sum_{s = \pm}^{} \sum_{l=0}^{\infty}\int\frac{d^2p}{(2\pi)^2}f(p_0,p_z;l,s).\label{eqII10}
\end{eqnarray} 
When combined with finite-temperature and finite-size effects, the Euclidean coordinate vectors must be rewritten with $x_0\in[0,\beta]$ and $x_z\in[0,L]$, $L$ being the size of the compactified spatial dimension. Then, taking the Matsubara prescription discussed in Eq.~(\ref{feynmanrule}) the integration over momenta space is modified accordingly, 
\begin{eqnarray}
\int\frac{d^4p}{(2\pi)^4}f(p) \rightarrow \frac{|Q_i |\omega }{2\pi} \frac{1}{\beta L } \sum_{s = \pm}^{}\sum_{l=0}^{\infty}  \sum_{ n_{\tau} ,n_z = - \infty}^{\infty} f \left( \tilde{\omega}_{n_{\tau}} , \bar{\omega}_{n_z} , |Q_i | \omega; l, s \right).
\label{feynmanrule2}
\end{eqnarray}

Hence, the thermodynamic potential and gap equations given by Eqs.~(\ref{effpot3}) and~(\ref{massa}) must be analyzed keeping in mind the thermodynamic relations and chiral condensates with the modified integration over momenta as in Eq.~(\ref{feynmanrule2}).
%In order to include magnetic effects on the system, we should remember that the fermionic propagator under external magnetic field in $z$ direction is given by Ritus propagator~\cite{Ritus,Emerson1}. In the euclidean space, in the limite  ${\mathbf{r}}^{\prime}\rightarrow \mathbf{r}$, its given by
%
%\begin{eqnarray}
%S_{E}(B) &\equiv& \left(\frac{\omega}{2\pi}\right)\sum_{\ell = 0}^{+\infty}\sum_{s=\pm 1}^{}\int \frac{d{p}_{\tau}}{(2\pi)}\frac{d{p}_{z}}{(2\pi)} \frac{(\slashed{\bar{p}}_{_E} - M_{i})}{\bar{p}_{_E}^{2}+M_{i}^{2}},
%\label{PropEucl}
%\end{eqnarray}
%where $\bar{p}_{_E}^{2}={p}_{\tau}^{2}+{p}_{z}^{2}+\omega (2 \ell +1 - s)$, $\omega \equiv Q_f B$ is the cyclotron frequency, $ s=\pm 1$ the spin of fermion and $\ell$ are the Landau levels. 
This means that the expressions for $\phi_i$ in Eq.~(\ref{phi2}) should be replaced by 
\begin{eqnarray}
\phi_{i} (T,L,\mu, \omega ) &=& -\frac{4N_{c} M_{i}|Q_i | \omega}{2\pi\beta L_{z}} \sum_{s = \pm}^{} \sum_{l = 0 }^{\infty} \int_{0}^{\infty} d\tau \exp[-\tau(M_{i}^{2}-\mu^{2})]\,\theta_{2}\left[\frac{2\pi\mu\tau}{\beta}\,;\,\exp(-4\pi^2 \tau/\beta^2)\right] \nonumber \\
&&\times \theta_{2}\left[0\,;\,\exp(-4\pi^2 \tau/L_{z}^2)\right] \exp\left[-\tau |Q_i |\omega \left(2 l + 1 - s \right)\right], 
\label{phi2mag}
\end{eqnarray}
which can be rewritten, after performing the proper time regularization procedure and the sum of geometrical series and spin polarization, as
\begin{eqnarray}
\phi_{i} (T,L,\mu, \omega ) &=& -\frac{4N_{c} M_{i} |Q_i | \omega}{2\pi\beta L_{z}} \int_{1/\Lambda^2}^{\infty} d\tau \exp[-\tau(M_{i}^{2}-\mu^{2})]\,\theta_{2}\left[\frac{2\pi\mu\tau}{\beta}\,;\,\exp(-4\pi^2 \tau/\beta^2)\right] \nonumber \\
&&\times \theta_{2}\left[0\,;\,\exp(-4\pi^2 \tau/L_{z}^2)\right] \left[ \frac{\exp\left(2 |Q_i | \omega \tau \right)}{-1+\exp\left(2 |Q_i | \omega \tau \right)}+  \frac{1}{-1+\exp\left(2 |Q_i | \omega \tau \right)} \right].
\label{phi3mag}
\end{eqnarray}

In the next section, we will discuss the $(T,L_{j},\mu, \omega ) $-dependence of thermodynamic quantities introduced above.

%%%%%%%%%%%%%%%%%%%%%%%%%%%%%%%%%%%%%%%%%%%%%%%%%%%%%%%%%%%%%%%%%%%%%%%%%%%%%%%%%%%%%%%%%%%%%%%%%%%%%%%%%%%%%%%%%%%
%%%%%%%%%%%%%%%%%%%%%%%%%%%%%%%%%%%%%%%%%%%%%%%%
\section{Phase structure}
%%%%%%%%%%%%%%%%%%%%%%%%%%%%%%%%%%%%%%%%%%%%%%%%%%%%%%%%%%%%%%%%%%%%%%%%%%%%%%%%%%%%%%%%%%%%%%%%%%%%%%%%%%%%%%%%%%
%%%%%%%%%%%%%%%%%%%%%%%%%%%%%%%%%%%%%%%%%%%%%%%%

We devote this section to the analysis of the phase structure of the system, focusing on how it behaves with the change of the relevant parameters of the model and, in special, the influence of the boundaries on the behavior of constituent quark masses $M_{u}$ and $M_{s}$, which are solutions of expressions given by Eq.~(\ref{massa});  explicitly, they are
\begin{eqnarray}
M_{u} & = & m_{u}-4G\phi_{u}+2\kappa\phi_{d}\phi_{s} , \nonumber \\
M_{d} & = & m_{d}-4G\phi_{d}+2\kappa\phi_{s} \phi_{u} , \nonumber 
\\
M_{s} & = & m_{s}-4G\phi_{s}+2\kappa \phi_{u} \phi_{d},
\label{Masses}
\end{eqnarray}
with $ M_{i} = M_{i}(T,L,\mu, \omega).$ We simplify the present study by fixing $L_{1}=L_{2}=L_{3}=L$, which means that the system consists in a $(u,d,s)$-quark gas constrained in a cubic box.  

%%%%%%%%%%%%%%%%%%%%%%%%%%%%%%%%%%%%%%%%%%%%%%%%%%%%%%%%%%%%%%%%%%%%%%%%%%%%%%%%%%%%%%%%%%%%%%%%%%%%%%%%%%%%%%%%%%%
%%%%%%%%%%%%%%%%%%%%%%%%%%%%%%%%%%%%%%%%%%%%%%%%
\subsection{Absence of magnetic field}
%%%%%%%%%%%%%%%%%%%%%%%%%%%%%%%%%%%%%%%%%%%%%%%%%%%%%%%%%%%%%%%%%%%%%%%%%%%%%%%%%%%%%%%%%%%%%%%%%%%%%%%%%%%%%%%%%%
%%%%%%%%%%%%%%%%%%%%%%%%%%%%%%%%%%%%%%%%%%%%%%%%

\begin{figure}
\centering
\includegraphics[{width=8.0cm}]{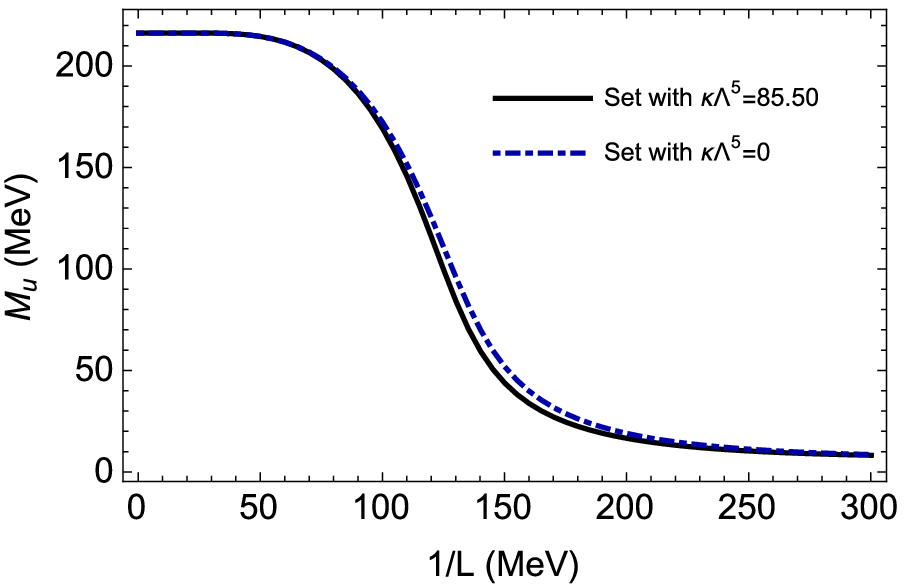} \\
\includegraphics[{width=8.0cm}]{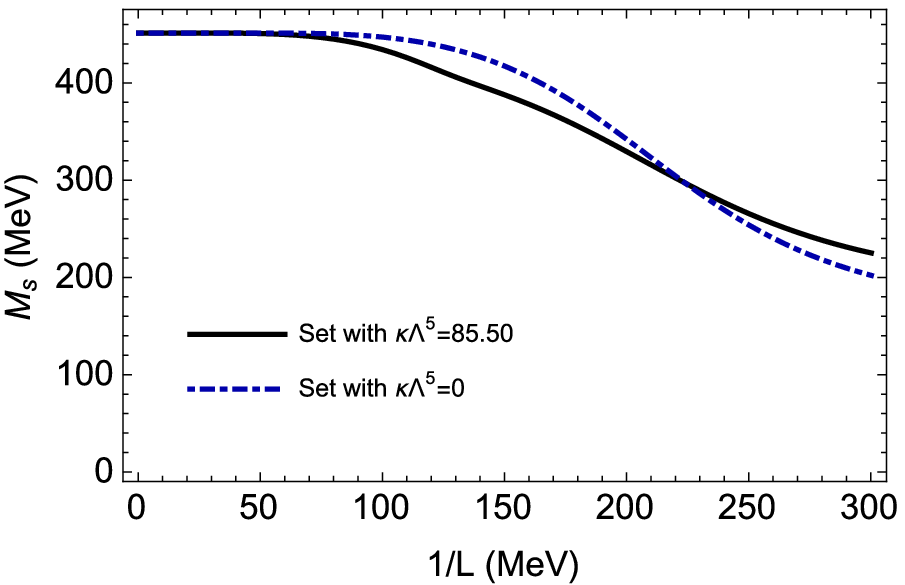} 
\caption{Constituent quark masses $M_u$ (top panel) and $M_s$ (bottom panel)  as functions of inverse of length $L$ at vanishing temperature and chemical potential. we have taken into account the situations with and without the 't Hooft interaction term (i.e. $\kappa \neq 0 $ or $\kappa = 0 $, respectively).}
\label{Massa1L1}
\end{figure}

Once the regularization procedure is chosen, one must determine the complete set of input  parameters that provides a satisfactory description of hadron properties at zero temperature and density. In this sense, we use the model parameters reported in Ref.~\cite{Kohyama:2016fif}, which have been estimated by the fitting in light of the following observables: $m_{\pi} = 138$ MeV, $f_{\pi} = 92$ MeV, $m_K = 495$ MeV and $m_{\eta^{\prime}} = 958$ MeV. Our choice of the set is the following: 
\begin{eqnarray}
	&&m_{u}  \approx  m_{d}=7.0\,\mathrm{MeV}\,;\, m_{s} = 195.6\,\mathrm{MeV} \,;\, \Lambda = 924.1\,\mathrm{MeV} \,;\, \nonumber \\
 &&G\Lambda^{2}   =  3.059\,;\, \kappa\Lambda^{5}=85.50.
	\label{parameters}   
\end{eqnarray}
The contribution of the  't Hooft interaction will be always taken into account unless explicitly stated otherwise, as in the case of Fig.~\ref{Massa1L1}, where the parameters  have been refitted in order to obtain the same dressed masses at zero temperature and density that correctly reproduce the relevant observables.
Moreover, since we work in the limit of isospin symmetry for current $u$ and $d$ quark masses, in absence of magnetic field the solutions of Eq.~(\ref{Masses}) satisfy $M_{u} \approx M_{d}$. Hence, throughout this subsection we will show and discuss the results keeping this fact in mind.

\begin{figure}
\centering
\includegraphics[{width=8.0cm}]{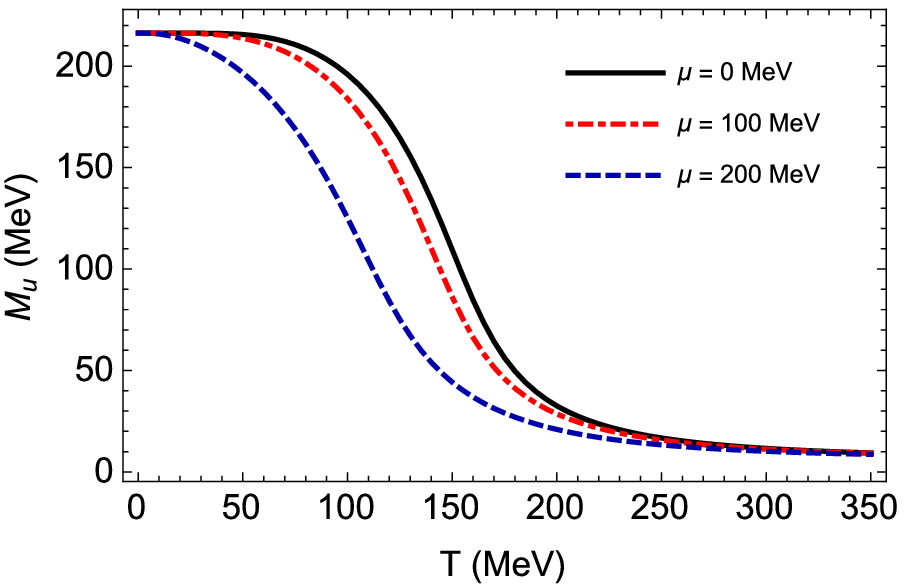} \\\includegraphics[{width=8.0cm}]{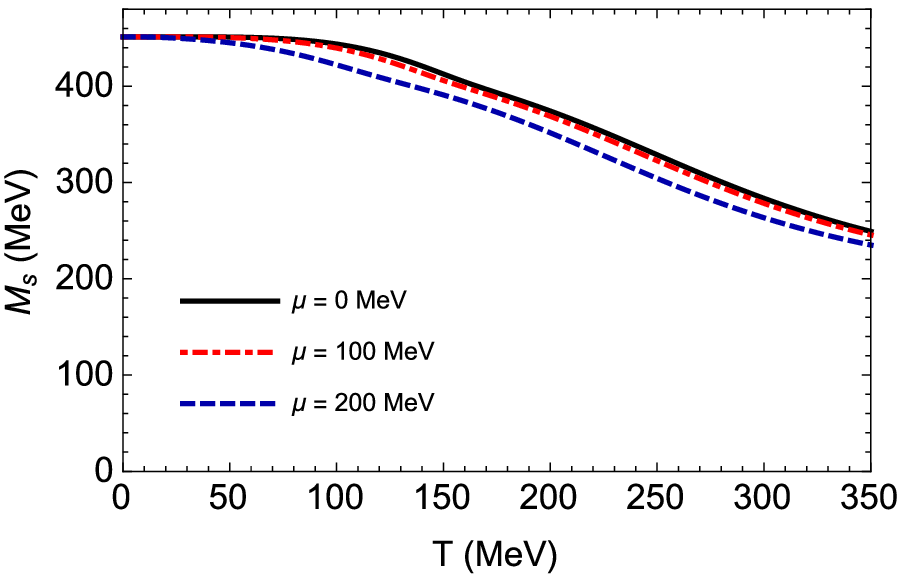}
\caption{Constituent quark masses $M_u$ (top panel) and $M_s$ (bottom panel) as functions of the temperature $T$, in the bulk limit $( L \rightarrow \infty )$ for different values of chemical potential $\mu$.}
\label{MassaBulk}
\end{figure}

We start, for completeness, by studying the behavior of the constituent quark masses under a change of parameters but without the presence of boundaries, which is basically the scenario described in Refs.~\cite{Buballa,Franceses}. 
In Fig.~\ref{MassaBulk} we have plotted  the values of $M_u$ (top panel) and $M_s$ (bottom panel) that are solutions of the gap equations in Eq.~ (\ref{Masses}) as functions of the temperature $T$, in the bulk limit $( L \rightarrow \infty )$ and taking different values of the chemical potential $\mu$.  It can be seen that the set of parameters given by Eq.~(\ref{parameters}) engenders the constituent quark masses $M_u \approx 216\,\mathrm{MeV} $ and $M_s \approx 451\,\mathrm{MeV} $ at vanishing temperature. 
In the range of smaller temperatures, there is no relevant modification. This phenomenon remains up to a certain temperature, where the masses start to decrease with the augmentation of $T$. Therefore the broken phase is inhibited  and a crossover transition takes place. Above a given temperature value, the dressed quark masses approach the magnitudes of the corresponding current quark masses, i.e. $m_{u} \approx 7.0\,\mathrm{MeV}\,;\, m_{s} \approx 195.6\,\mathrm{MeV}$. It is also worth mentioning that the constituent mass for the $s $-quark has a smoother drop than the one for the $u$-quark, falling to the $m_{s}$ magnitude at larger $T$. Another feature is that at higher temperatures the system tends faster toward the symmetric chiral phase as the chemical potential increases.

We now investigate in detail the influence of boundaries on the phase structure of the system, considering the $(u,d,s)$-quark gas in a region delimited by a cubic box. In Fig.~\ref{Massa1L1} we have plotted the values of $M_{i}$ that are solutions of the gap equations in Eq. (\ref{Masses}) as function of the inverse length $1/L$, at vanishing temperature and chemical potential. For the sake of comparison with the existing literature (e.g., Refs.~\cite{Buballa,Franceses,Wang:2018qyq}), we have taken into account the situations with and without the 't Hooft interaction term (i.e., $\kappa \neq 0 $ or $\kappa = 0 $, respectively). 
Note that in the case of a vanishing $\kappa$, the parameters in Eq~(\ref{parameters}) have been readjusted  keeping the values of $\Lambda$ and $m_u$ fixed and varying $G$ and $m_s$ to obtain the same dressed masses estimated at zero temperature and density for the situation with $\kappa \neq 0$. Then, in this context we have worked with the set, 
\ben
& & m_{u}  \approx  m_{d} = 7.0\,\mathrm{MeV}\,;\, m_{s} = 167.8\,\mathrm{MeV} \,;\, \Lambda = 924.1\,\mathrm{MeV} \,;\, \nonumber \\
& & G\Lambda^{2}   =  3.900\,;\, \kappa = 0. 
\label{parameters2}
\een
We can observe that the bulk approach appears as a good approximation in the range of greater values of $L$ (up to $\approx 3-4 $~fm). 
%It can be remarked that the presence of the 't Hooft interaction, which engenders a system of coupled gap equations shown in Eq.~(\ref{Masses}), modifies the constituent quark masses in the bulk, yielding larger values: $M_u$ increases from $263.4$ to $312.3$~MeV, while $M_s$ grows from $387.1$ to $425.4$~MeV. We can observe that the bulk approach appears as a good approximation in the range of greater values of $L$ (up to $\approx 3 $~fm). 
The dressed masses are affected by the presence of boundaries as the size of the system decreases, acquiring smaller values smoothly. In this region of lower $L$,  we see that the presence of the 't Hooft interaction, which engenders a system of coupled gap equations shown in Eq.~(\ref{Masses}), keeps the essential features unchanged when the parameter set is properly chosen to reproduce the masses $M_i$ in the bulk. Only in the case of $M_s$ the contributions coming from $ \kappa \neq 0$ yield a different decreasing more pronounced than the situation of a vanishing $\kappa$. Also, we notice that below a given value of $L$, the dressed quark masses tend to the corresponding current quark masses. In particular, this happens at $L \approx 1$-$1.5$~fm for $ u,d$-quarks. So, analogously to the behavior described in the previous figures, $M_s$ has a smoother falloff to the current quark mass than $M_u$. Thus, the  dependence of the phase structure of the system on the inverse length $1/L$ is qualitatively similar to the one found for the temperature, which is expected, due to the equivalent nature between $1/L $ and $T$, manifested in the generalized Matsubara prescription in Eq~(\ref{Matsubara}).  

\begin{figure}
\centering
\includegraphics[{width=8.0cm}]{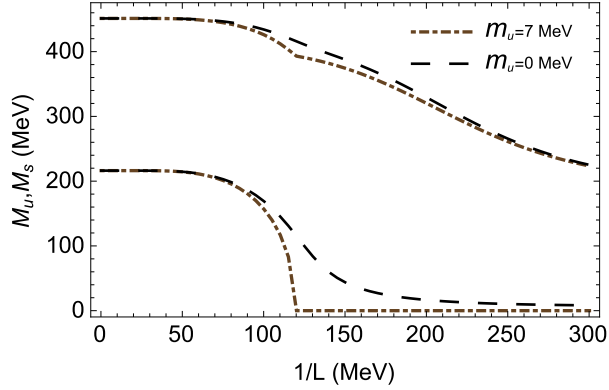}
\caption{Constituent quark masses $M_u$ and $M_s$ as functions of the inverse length $1/L$ at vanishing temperature and chemical potential, taking the chiral limit for the $u,d$ quarks, i.e., setting $m_u = 0$.}
\label{Massa1L2}
\end{figure}

The plot in Fig.~\ref{Massa1L2} is the same as in Fig.~\ref{Massa1L1} for the case $\kappa \neq 0 $, but with the solutions for $M_u$ and $M_s$ obtained at vanishing temperature and chemical potential taking the chiral limit for the $u,d$ quarks, i.e., setting $m_u = 0$. Once more, as in the previous situation, in the chiral limit we have  refitted the parameters to give a good prescription of the hadronic observables at zero temperature and density and in the bulk. 
Therefore, keeping the value of $\Lambda$ and $m_s$ fixed and varying $G$ and $\kappa$ to obtain the same dressed masses estimated at $T = \mu = 1/L = 0$, we have employed the set 
\ben
& & m_{u}  \approx  m_{d}=0\,\mathrm{MeV}\,;\, m_{s} = 195.6 \,\mathrm{MeV} \,;\, \Lambda = 924.1\,\mathrm{MeV} \,;\, \nonumber \\
& &  G\Lambda^{2}   =  2.903 \,;\,  \kappa \Lambda^{5} = 114.64.
\label{parameters3}
\een
Again, the restoration of chiral symmetry is induced at smaller sizes of the system, lost in the bulk. In this context, the results suggest the existence of a critical volume $V_c = (L_c )^3 \approx (1.5)^3-(2.0)^3 \, \mathrm{fm}^3 $ at which the constituent mass for $u$-quark reaches zero value via a second-order phase transition. On the other hand,  $M_s $ goes down to the value of $m_{s}$ softly.

%
%{\color{red} In the Fig.~$1$ we show the behaviour of the effective masse $M_u$ with respect to temperature in bulk form, in the range $0$~MeV $  \lesssim T  \lesssim 400$~MeV, that is, we used the Eqs.~(\ref{phibulk}) and (\ref{massa}).

%In the Figs.~$3$ and $4$, we show the finite size effects on the model, in this case, the system has cubic shape with size within range $0.65$~fm~$ \lesssim L \lesssim $ $400$~fm. Taking into account the conversion factor $1$~MeV$^{-1} = 196.9$~fm, this finite range is equivalente to $303$~MeV~$\gtrsim 1/L \gtrsim 0.5$~MeV.

\begin{figure}
\centering
\includegraphics[{width=8.0cm}]{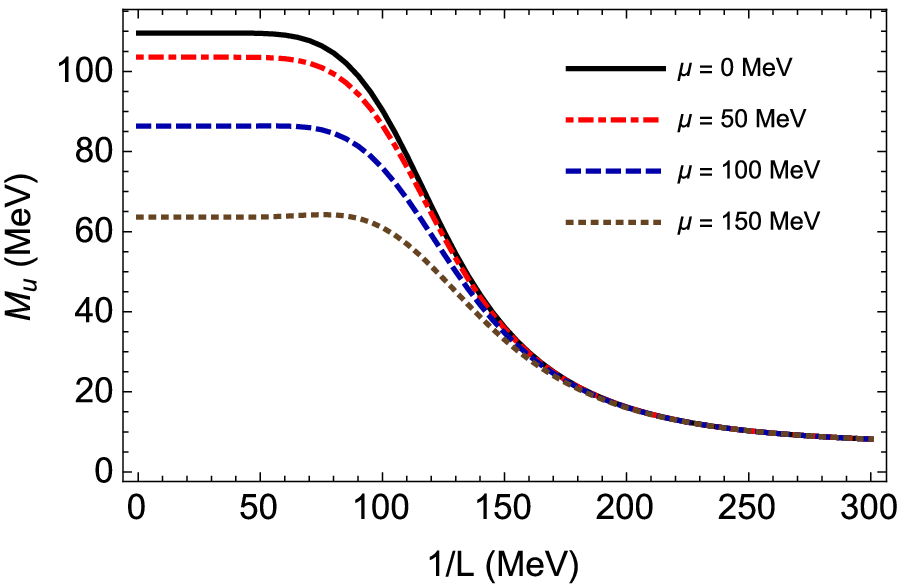} \\\includegraphics[{width=8.0cm}]{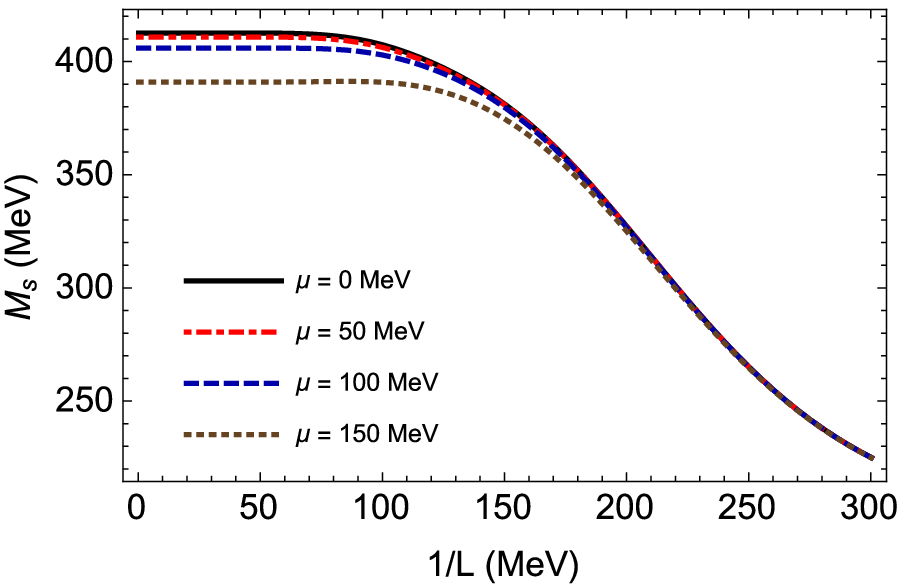}
\caption{ Constituent quark masses $M_u$ (top panel) and $M_s$ (bottom panel) as functions of the inverse length $1/L$ at $T= 150$ MeV, for different values of the chemical potential $\mu$. }
\label{MassaFiniteL}
\end{figure}

Let us concentrate at this point on the evaluation of the combined effects of boundaries, finite temperature and finite chemical potential on the system. They can be better described from Fig.~\ref{MassaFiniteL}, in which 
are plotted the values of effective masses that are solutions of the gap equation in Eq. (\ref{Masses}) as function of inverse length $x=1/L$ for different values of  $\mu$,  keeping the temperature fixed at 150~MeV. As suggested in previous figures,
in the bulk ($1/L \rightarrow 0$ or $L\rightarrow\infty$), $M_{u}$ and $M_{s}$ suffer a decrease as the chemical potential grows, and diminishes even more with the drop of $L$ and tend to the respective current quark masses.  Also, at smaller $L$ the dressed masses associated to different values of $\mu$ converge to the values of corresponding current quark masses.

\begin{figure}
\centering
\includegraphics[{width=8.0cm}]{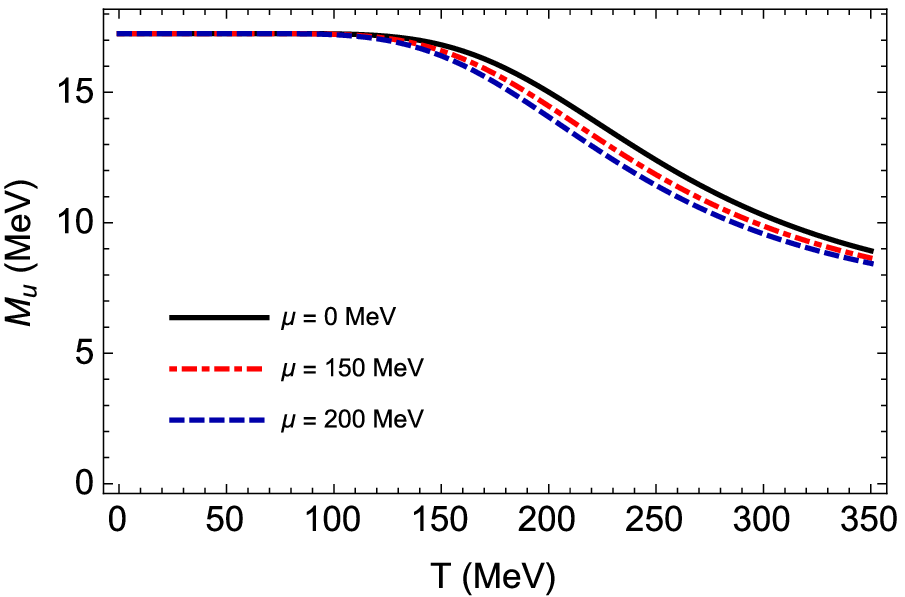} \\
\includegraphics[{width=8.0cm}]{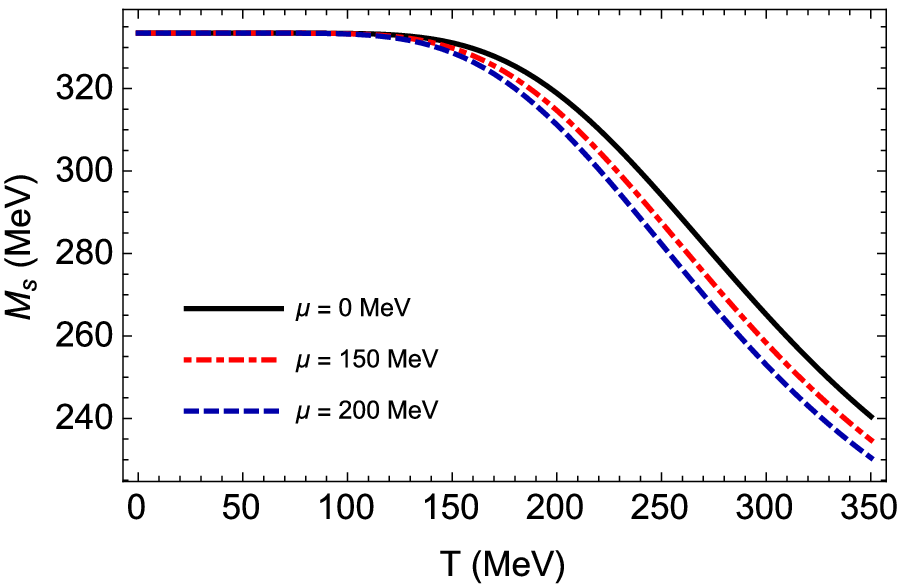}
\caption{Constituent quark masses $M_u$ (top panel) and $M_s$ (bottom panel) as functions of temperature $T$  at $L = 1$~fm, taking different values of chemical potential $\mu$.}
\label{MassaL1fm}
\end{figure}

\begin{figure}
\centering
\includegraphics[{width=8.0cm}]{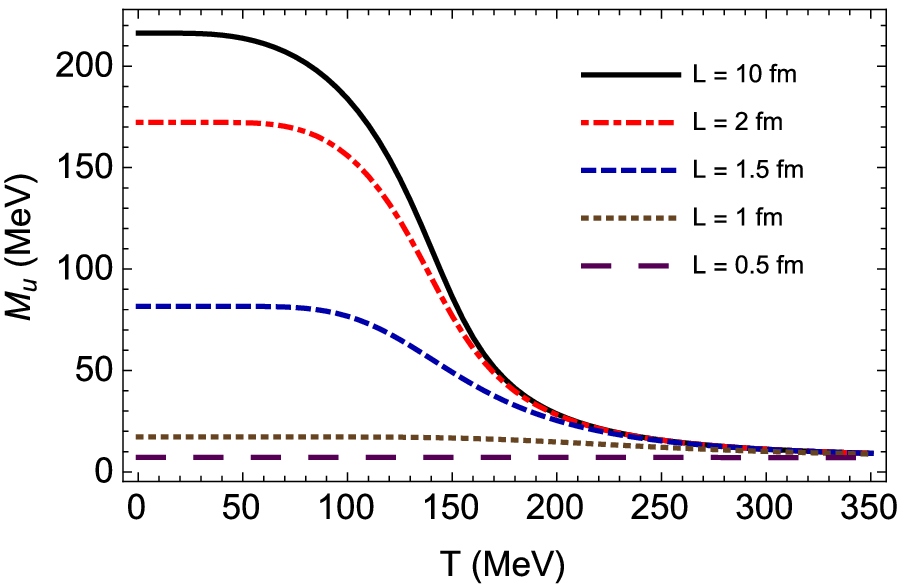} \\
\includegraphics[{width=8.0cm}]{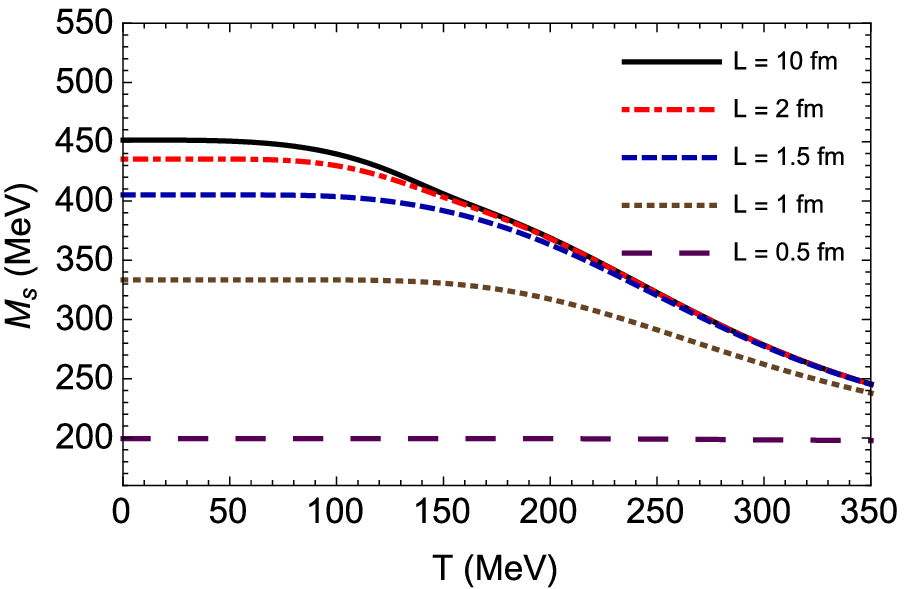}
\caption{Constituent quark masses $M_u$ (top panel) and $M_s$ (bottom panel) as functions of the temperature $T$  at $\mu = 100$~MeV, taking different values of size $L$.}
\label{MassavariosL}
\end{figure}

This analysis is completed with Figs.~\ref{MassaL1fm} and \ref{MassavariosL}, where are  plotted  the values of $M_u$ and $M_s$ as functions of the temperature $T$, taking different values of chemical potential $\mu$ and size $L$, respectively. When compared with Fig.~\ref{MassaBulk}, the plots in Fig.~\ref{MassaL1fm} manifest the smaller values of constituent quark masses reached in the case of finite volume. Fig.~\ref{MassavariosL} shows the values of the size of cubic box in which $M_u$ and $M_s$ stand with the corresponding values of current quarks masses in all range of temperature.  Hence, our findings suggest that the presence of boundaries disfavors the maintenance of long-range correlations, inducing the inhibition of the broken phase.

%%%%%%%%%%%%%%%%%%%%%%%%%%%%%%%%%%%%%%%%%%%%%%%%%%%%%%%%%%%%%%%%%%%%%%%%%%%%%%%%%%%%%%%%
%%%%%%%%%%%%%%%%%%%%%%%%%%%%%%%%%%%%%%%%%%%%%%%%%%%%%%%%%%%%%%%%%%%%%%%%%%%%%%%%%%%%%%%%
\subsection{Magnetic background effects}
%%%%%%%%%%%%%%%%%%%%%%%%%%%%%%%%%%%%%%%%%%%%%%%%%%%%%%%%%%%%%%%%%%%%%%%%%%%%%%%%%%%%%%%%
%%%%%%%%%%%%%%%%%%%%%%%%%%%%%%%%%%%%%%%%%%%%%%%%%%%%%%%%%%%%%%%%%%%%%%%%%%%%%%%%%%%%%%%%

Now we investigate the combined effects of finite volume and magnetic background on the phase structure of the thermal gas of quark matter, taking into account that here we have only one spatial compactified coordinate. We remember that in gap equations (\ref{Masses}) of present situation we use a different expression for the condensates, i.e. Eq.~(\ref{phi3mag}) rather than~(\ref{phi2}), in which a dimensional reduction of four-momentum integrals in Feynman rules took place. The parameter set used is shown in Eq.~(\ref{parameters}), which gives the correct hadronic observables at $T, \mu, 1/L, \omega = 0$.
 
Furthermore, noticing that the coupling of the quarks to the electromagnetic field depends on the quark flavor (see discussion in subsection~\ref{magnetic-field}), then in principle we cannot adopt the solutions of Eq.~(\ref{Masses}) satisfying $M_{u} \approx M_{d}$, as done previously. That being so, we solve the system of three gap equations in~(\ref{Masses}) to assess the differences between  $M_{u} $ and $ M_{d}$ engendered by the coupling to magnetic background.

\begin{figure}
	\centering
	\includegraphics[{width=8.0cm}]{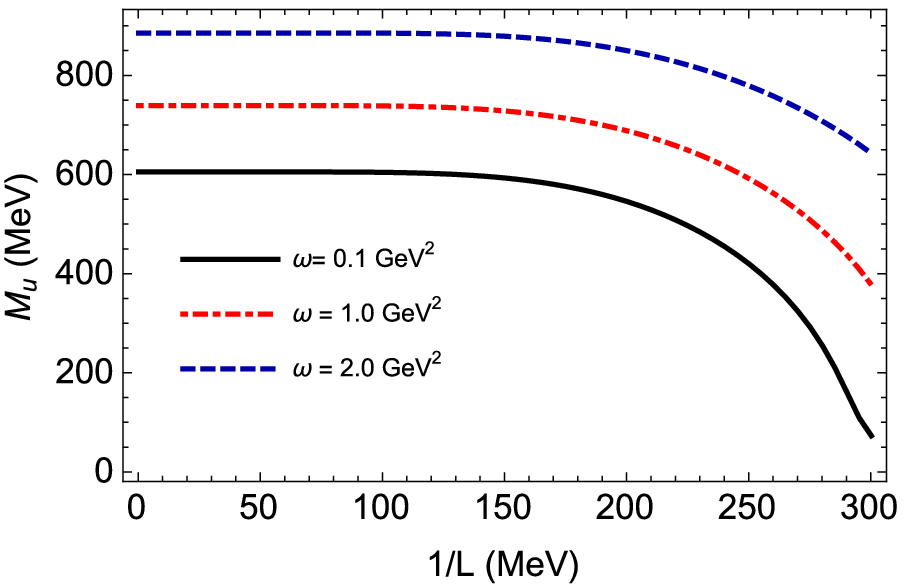} \\
	\includegraphics[{width=8.0cm}]{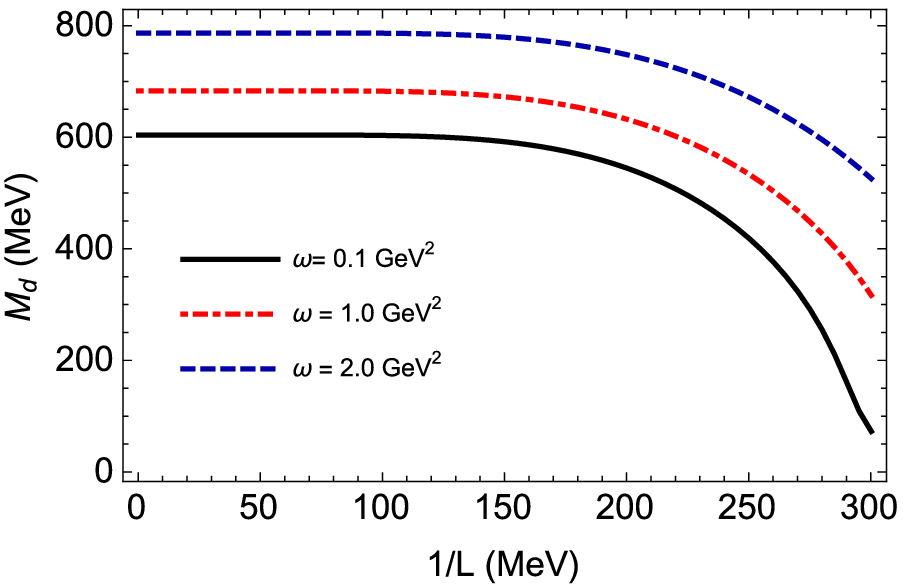} \\
	\includegraphics[{width=8.0cm}]{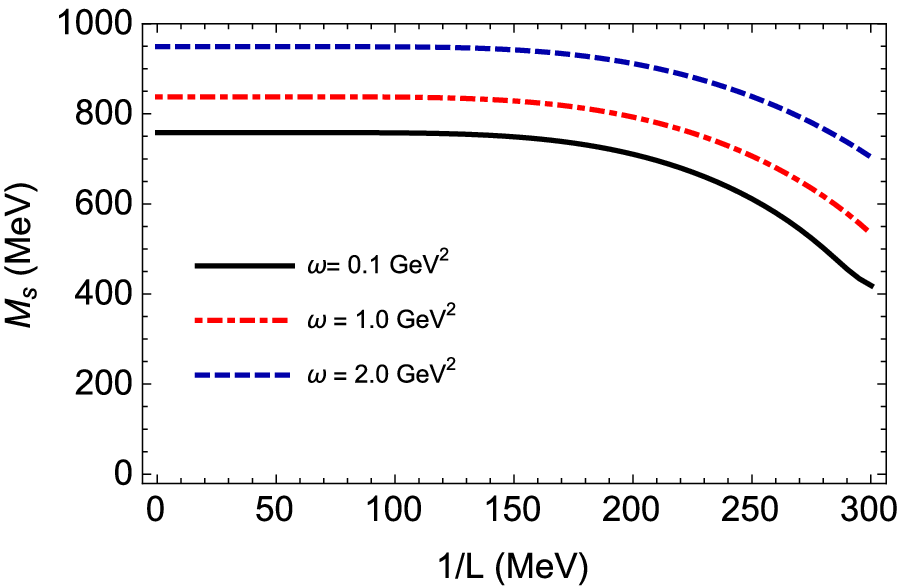}
	\caption{ Constituent quark masses $M_u$ (top panel), $M_d$ (middle panel) and $M_s$ (bottom panel) as functions of the inverse of length $1/L$  at $T, \mu = 0$~MeV, taking different values of external magnetic field.}
	\label{fig:Mag1}
\end{figure}

In Fig.~\ref{fig:Mag1} is plotted the values of constituent quark masses as functions of inverse of thickness $x=1/L$ for different values of  $\omega $,  keeping temperature and chemical potential fixed. For higher values of the considered range of magnetic field strength,  we remarked  that the approximate isospin symmetry for the constituent masses of the $u$ and $d$ quarks is broken, with $M_u$ being bigger than $M_d$ due to the larger coupling of $u$ quark to magnetic background.
Also, we see that in the bulk ($x\rightarrow 0$ or $L\rightarrow\infty$), the presence of an external magnetic field engenders the augmentation of $M_{i}$. In other words, the raise of field strength enhances the broken phase. This feature is  expected and called in literature as magnetic catalysis effect. This phenomenon remains in all considered range of $L$. On the other hand, as previously the reduction of $L$ engenders a reduction of constituent masses, but in a more slightly way than the situation without magnetic field. Also, in the range of smaller values of $L$ the dressed masses associated to different values of $\omega$ converge to the values of corresponding current quark masses. 
The interesting point is that the growth of field strength induces smaller values for $L$ at which the system stand with the values $m_i$.  Thus, the combination of finite size and magnetic effects generates a competition between then, since the former inhibits the broken phase whereas the last one yields its stimulation.

\begin{figure}
	\centering
	\includegraphics[{width=8.0cm}]{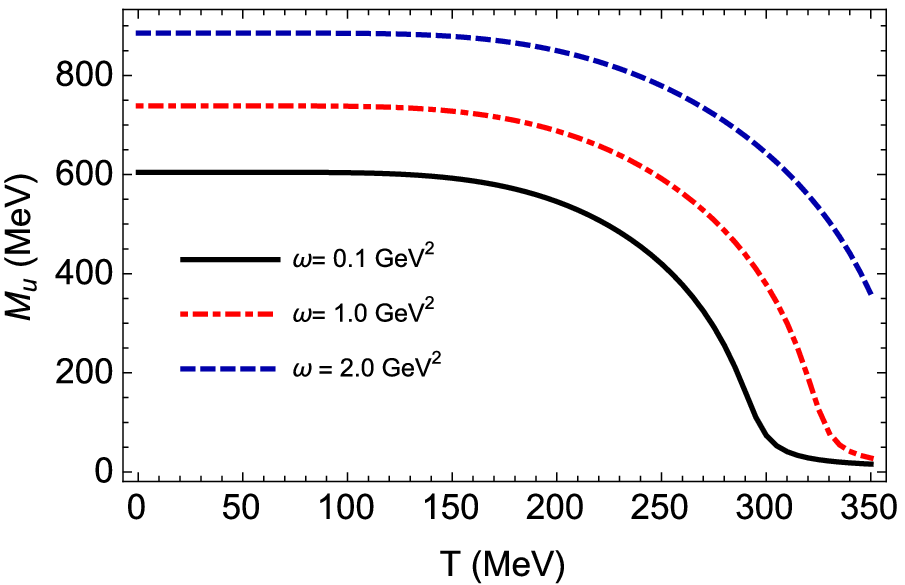} \\
	\includegraphics[{width=8.0cm}]{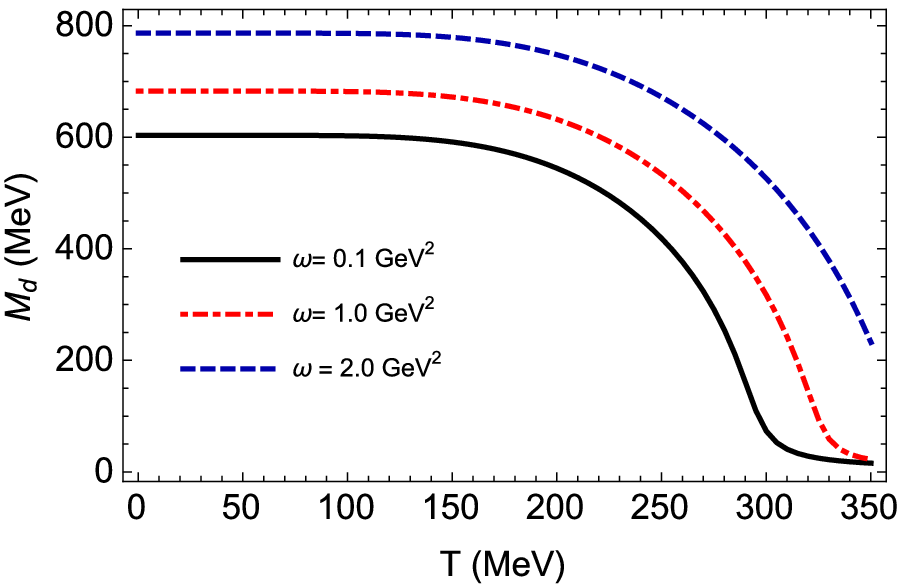} \\
	\includegraphics[{width=8.0cm}]{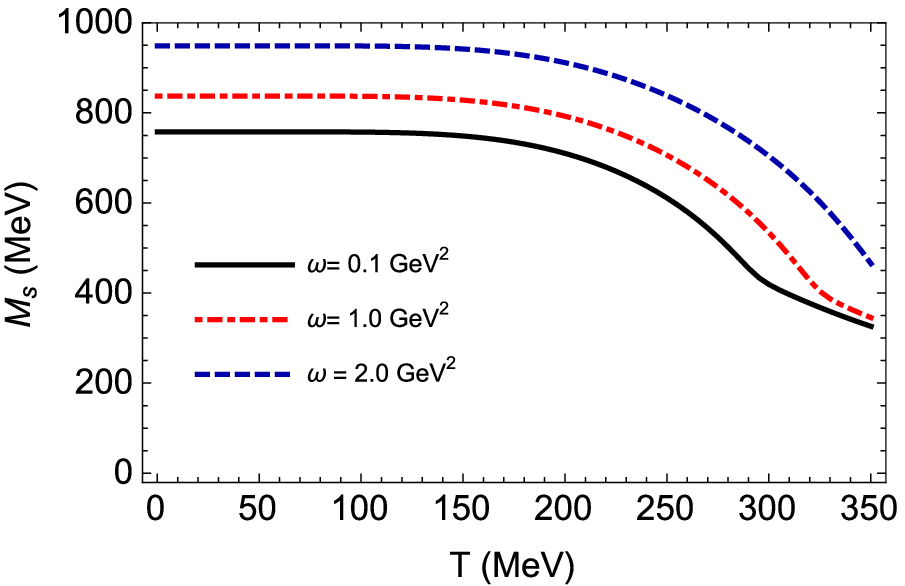}
	\caption{ Constituent quark masses $M_u$ (top panel), $M_d$ (middle panel) and $M_s$ (bottom panel) as functions of the Temperature $T$ at $\mu = 0$~MeV, taking different values of external magnetic field for the finite size in the $z$ fixed at $L = 2$~fm.}
	\label{fig:Mag2}
\end{figure}

\begin{figure}
	\centering
	\includegraphics[{width=8.0cm}]{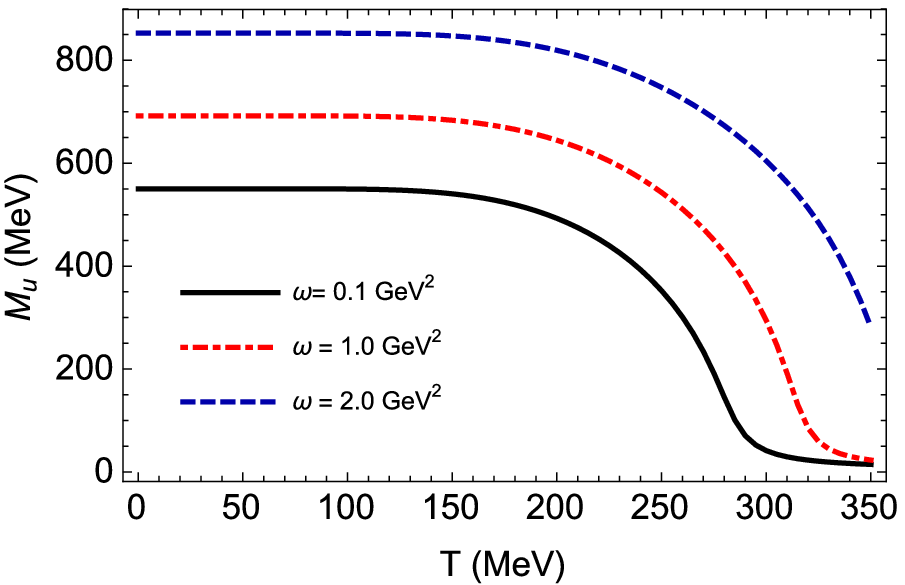} \\
	\includegraphics[{width=8.0cm}]{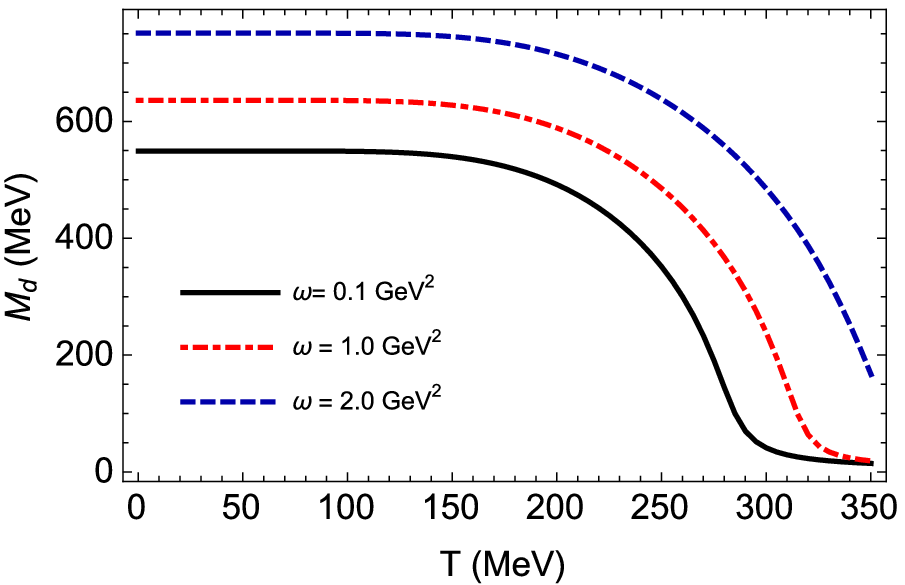} \\
	\includegraphics[{width=8.0cm}]{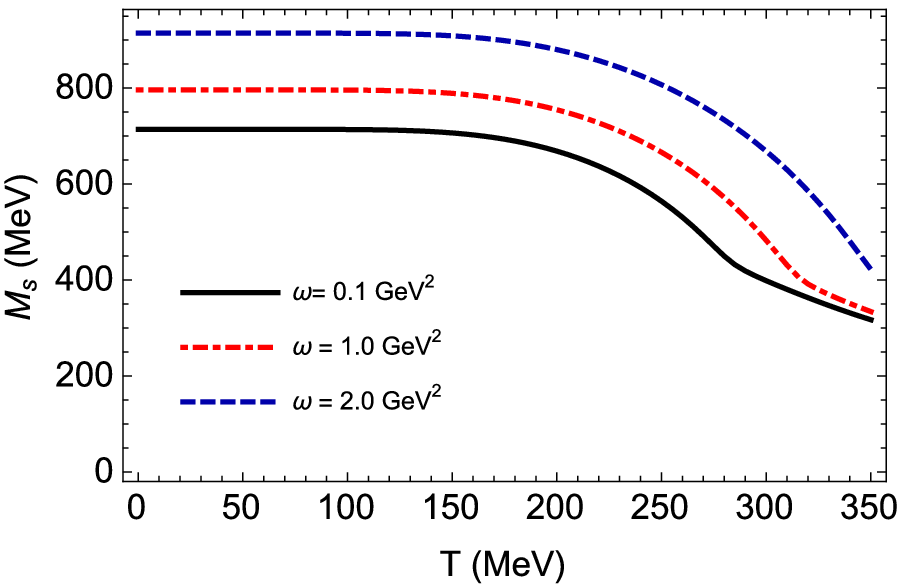}
	\caption{ Constituent quark masses $M_u$ (top panel), $M_d$ (middle panel) and $M_s$ (bottom panel) as functions of the Temperature $T$ at $\mu = 0$~MeV, taking different values of external magnetic field for the finite size in the $z$ fixed at $L = 1$~fm. }
	\label{fig:Mag3}
\end{figure}

We complement this analysis with Figs.~\ref{fig:Mag2} and \ref{fig:Mag3}, where are  plotted  the values of $M_i$  as functions of the temperature $T$, taking different values of field strength $\omega$ and size $L$, respectively, which can be compared with Figs.~\ref{MassaL1fm} and~\ref{MassavariosL}. Again, they manifest the smaller values of constituent quark masses reached in the case of finite size and greater temperatures. Since the magnetic field increases the constituent quark masses, the critical size $L_c$ at which $M_i$ stand with the corresponding values of current quarks masses in all range of temperature should be even smaller than the situation without magnetic field (0.5-1.0 fm).

%%%%%%%%%%%%%%%%%%%%%%%%%%%%%%%%%%%%%%%%%%%%%%%%%%%%%%%%%%%%%%%%%%%%%%%%%%%%%%%%%%%%%%%%

We conclude this section by discussing some issues. We notice that the relevance of the  regularization procedure and parametrization in effective models can not be underestimated. The findings obtained above are naturally dependent on the set of parameters considered as input in Eq.~(\ref{Masses}). Modifications in these choices will obviously alter the magnitudes of the constituent quark masses and the ranges of $(T,L,\mu, \omega)$ which engender changes on their behavior, making them drop faster or slower and  accordingly reshape the curves plotted in previous figures. Thus, since in the present approach we have chosen the set of input parameters that provides a satisfactory description of hadron properties at $T, \mu , 1/L, \omega = 0$ (according to Ref.~\cite{Kohyama:2016fif}), the comparison with existing literature only makes sense with those works that have proceeded in the same way.

Notwithstanding the points raised above, we stress the main results of this work as follows. In the end, we see that the phase structure of the system is strongly affected by the combined variation of relevant variables, depending on  the competition among their respective effects. At smaller temperatures, with the system in the broken phase, the bulk approach seems a good approximation in the range of greater values of the size of the system $L$, since the constituent masses do not suffer modification.  Nevertheless, keeping $T$ fixed, the reduction of $L$ from a given value engenders a reduction of $M_i$  via a crossover phase transition to their corresponding current quark masses. Yet at the same temperature and at the range $L < L_c$, the broken phase is disfavored due to both increasing of chemical potential and the drop of $L$. On the other hand, when a magnetic background is present the broken phase is stimulated, $M_i$ acquires greater values and the increase of field strength induces even smaller values for $L$ to reach symmetric phase.  

Finally, we should point out the combined effect of an external magnetic field and the choice of boundary conditions in this model. The boundary effect in the context of antiperiodic condition has been studied in detail and summarized in previous paragraph. Then, we deserve some comments to the contribution of the compactified dimension length in the situation of periodic boundary conditions. After the compactification of the $z$-coordinate in the energy eigenvalues in Eq.~(\ref{eqII9}),  it can be seen that for the periodic condition, we have the prescription $ p_z \rightarrow \bar{\omega} _{n_z} \equiv 2\pi n_{z} /L_z  $, and no energy gap is yielded between the vacuum and the lowest Landau level $l=0$ at zero momentum $n=0$ in the chiral limit. This makes possible the vacuum condensation of fermion pairs in the infrared range of spectrum (see Ref.~\cite{Ferrer:1999gs} for a detailed discussion). In this sense, the presence of magnetic field catalyzes the dynamical breaking of chiral symmetry.
Another perspective to see this is looking at the chiral condensate in Eq.~(\ref{phi3mag}) and gap equations in~(\ref{Masses}), in which the Jacobi $\theta$-function $\theta_2$ must be replaced by the $\theta_3$ in the scenario of periodic condition. Thanks to the different behaviors of the $\theta$-functions, the consequence of this replacement is relevant: while in the antiperiodic case the magnetic catalysis is inhibited, with the length $ L $ playing a role similar to the inverse of temperature $\beta$, in the periodic situation the boundary has the opposite effect: symmetry breaking is enhanced, and accordingly the constituent quark masses $M_i$ acquire higher values as $L$ diminishes. Thus, there is not a critical value of the size $L_c$ in which the symmetry is restored, and the combined effect of magnetic background and boundary conditions in periodic case strengthens the broken phase. 

We must emphasize, however, that in our understanding it seems more appropriate to adopt the same boundary condition in both spatial and temporal directions, as remarked in subsection~\ref{Gen-Matsubara} and supported by a large number of studies using effective models~\cite{Gasser:1986vb,Klein:2017shl,Shi:2018swj,Wang:2018qyq}.

%%%%%%%%%%%%%%%%%%%%%%%%%%%%%%%%%%%%%%%%%%%%%%%%%%%%%%%%%%%%%%%%%%%%%%%%%%%%%%%%%%%%%%%%%%%%%%%%%%%%%%%%%%%%%%%%%%%%%%%%%%%%%%%%%%%%%%%%%%%%%%%%%%%%%%%%%%%%%%%%%%%%%%%%%%%%%%%%%%%%%%%%%%%%%%%%%%%%%%%%%%%%%%%%%%%%%%%%%%%%%%%%%%%%%%%%%%%%%%%%%%%%%%%%%%%%%%%%%%%%%%%%%%%%%%%%%%%%%%%%%%%%
%
\section{Concluding Remarks}
%%%%%%%%%%%%%%%%%%%%%%%%%%%%%%%%%%%%%%%%%%%%%%%%%%%%%%%%%%%%%%%%%%%%%%%%%%%%%%%%%%%%%%%%%%%%%%%%%%%%%%%%%%%%%%%%%%%%%%%%%%%%%%%%%%%%%%%%%%%%%%%%%%%%%%%%%%%%%%%%%%%%%%%%%%%%%%%%%%%%%%%%%%%%%%%%%%%%%%%%%%%%%%%%%%%%%%%%%%%%%%%%%%%%%%%%%%%%%%%%%%%%%%%%%%%%%%%%%%%%%%%%%%%%%%%%%%%%%%%%%%%%
%

In this work we have analyzed the finite volume and magnetic effects on the phase structure of a generalized version of the Nambu--Jona-Lasinio model with three quark flavors. By making use of mean-field approximation and the Schwinger proper time method in a toroidal topology with antiperiodic conditions, we have investigated the gap equation solutions under the change of the size of compactified coordinates, temperature, chemical potential and strength of external magnetic field.
The thermodynamic behavior is strongly affected by the combined effects of relevant variables, depending on the range of their change.  

Looking at the influence of a finite volume, we can remark the main results of this work: at smaller temperatures, with the system in the broken phase, the bulk approach seems a good approximation in the range of greater values of the volume of the system $V = L^3$, since the constituent masses are not modified. Nevertheless, while keeping $T$ fixed, there is a range in which the reduction of $L$ (remarkably at $L\approx 0.5 - 3 $~fm) engenders a reduction of the constituent masses for $u,d,s$ quarks through a crossover phase transition to the their corresponding current quark masses. After all, the broken phase is disfavored due to both increasing of temperature and chemical potential, and the drop of $L$. However, in the region in which the broken phase is inhibited, the current quark masses are not affected by the change of variables.

%%% 
Looking at the magnetic background influence, only phenomena associated to the magnetic catalysis take place: an enhancement of broken phase occurs in all considered ranges of temperature and size. When a magnetic background is present, constituent quark masses acquire greater values and the increase of field strength induces smaller values for $L$ and greater values of $T$ at which the $M_i$ tend to the respective values of current quark masses. For higher values of magnetic field strength, we remarked  that the approximate isospin symmetry for the constituent masses of the $u$ and $d$ quarks is no longer valid, with $M_u$ being bigger than $M_d$ due to the larger coupling of $u$ quark to magnetic background.
%%%

Finally, we emphasize that the results outlined above can give us insights about the finite-volume and magnetic effects on the critical behavior of quark matter produced in heavy-ion collisions. In particular, estimations have been done on the range of size of the system at which the bulk approximation looks a reasonable one. Further studies are necessary in order to verify the efficacy of the proposed framework.

\acknowledgments
%
%EBSC would like thanks ICE/UNIFESSPA by partial release to work in this paper and Erich Cavalcante for his help with the graphics.

This paper is dedicated to the memory of Edson B. S. Corr\^{e}a. The authors would like to thank the Brazilian funding agencies CNPq, CAPES. 
L.M.A. would like to thank the Brazilian funding agencies CNPq (contracts 308088/2017-4 and 400546/2016-7) and FAPESB (contract INT0007/2016) for financial support. E.B.S.C. would like to thank E. Cavalcanti for helpful discussions.

%%%%%%%%%%%%%%%%%%%%%%%%%%%%%%%%%%%%%%%%%%%%%%%%%%%%%%%%%%%%%%%%%%%%%%%%%%%%%%%%%%%%%%%%%%%%%%%%%%%%%%%%%%%%%%%%%%%%%%%%%%%%%%%%%%%%%%%%%%%%%%%%%%%%%%%%%%%%%%%%%%%%%%%%%%%%%%%%%%%%%%
%


\begin{thebibliography}{99}

\bibitem{qgpdisc}      J.~Adams {\it et al.} [STAR Collaboration],
                       %``Experimental and theoretical challenges in 
                       %the search for the quark gluon plasma: 
                       %The STAR Collaboration's critical assessment 
                       %of the evidence from RHIC collisions,''
                       Nucl.\ Phys.\ A {\bf 757}, 102 (2005).

\bibitem{rev-qgp}      P. Braun-Munzinger, V. Koch,
                       T. Schafer, and J. Stachel,
                       Phys. Rep. {\bf 621}, 76 (2016).


\bibitem{Prino:2016cni} 
  F.~Prino and R.~Rapp,
  %``Open Heavy Flavor in QCD Matter and in Nuclear Collisions,''
  J.\ Phys.\ G {\bf 43}, no. 9, 093002 (2016).
%  doi:10.1088/0954-3899/43/9/093002
%  [arXiv:1603.00529 [nucl-ex]].
  %%CITATION = doi:10.1088/0954-3899/43/9/093002;%%
  %55 citations counted in INSPIRE as of 29 Dec 2018



%\cite{Pasechnik:2016wkt}
\bibitem{Pasechnik:2016wkt} 
  R.~Pasechnik and M.~Šumbera,
  %``Phenomenological Review on Quark–Gluon Plasma: Concepts vs. Observations,''
  Universe {\bf 3}, no. 1, 7 (2017).
 % doi:10.3390/universe3010007
 % [arXiv:1611.01533 [hep-ph]].
  %%CITATION = doi:10.3390/universe3010007;%%
  %18 citations counted in INSPIRE as of 29 Dec 2018
 
 
%\cite{Bass:1998qm}
\bibitem{Bass:1998qm} 
  S.~A.~Bass {\it et al.},
  %``Reaction dynamics in Pb + Pb at the CERN / SPS: From partonic degrees of freedom to freezeout,''
  Prog.\ Part.\ Nucl.\ Phys.\  {\bf 42}, 313 (1999). 
%  doi:10.1016/S0146-6410(99)00086-1
%  [nucl-th/9810077].
  %%CITATION = doi:10.1016/S0146-6410(99)00086-1;%%
  %19 citations counted in INSPIRE as of 29 Dec 2018
   
%\cite{Palhares:2009tf}
\bibitem{Palhares:2009tf} 
  L.~F.~Palhares, E.~S.~Fraga and T.~Kodama,
  %``Chiral transition in a finite system and possible use of finite size scaling in relativistic heavy ion collisions,''
  J.\ Phys.\ G {\bf 38}, 085101 (2011). 
%  doi:10.1088/0954-3899/38/8/085101
%  [arXiv:0904.4830 [nucl-th]].
  %%CITATION = doi:10.1088/0954-3899/38/8/085101;%%
  %51 citations counted in INSPIRE as of 29 Dec 2018
  
  
%\cite{Graef:2012sh}
\bibitem{Graef:2012sh} 
  G.~Graf, M.~Bleicher and Q.~Li,
  %``Examination of scaling of Hanbury-Brown--Twiss radii with charged particle multiplicity,''
  Phys.\ Rev.\ C {\bf 85}, 044901 (2012).
%  doi:10.1103/PhysRevC.85.044901
%  [arXiv:1203.4071 [nucl-th]].
%  CITATION = doi:10.1103/PhysRevC.85.044901;%%
  %16 citations counted in INSPIRE as of 29 Dec 2018
  
  
%\cite{Luecker:2009bs}
\bibitem{Luecker:2009bs} 
  J.~Luecker, C.~S.~Fischer and R.~Williams,
  %``Volume behaviour of quark condensate, pion mass and decay constant from Dyson-Schwinger equations,''
  Phys.\ Rev.\ D {\bf 81}, 094005 (2010).
%  doi:10.1103/PhysRevD.81.094005
%  [arXiv:0912.3686 [hep-ph]].
  %%CITATION = doi:10.1103/PhysRevD.81.094005;%%
  %34 citations counted in INSPIRE as of 29 Dec 2018
  
%\cite{Li:2017zny}
\bibitem{Li:2017zny} 
  B.~L.~Li, Z.~F.~Cui, B.~W.~Zhou, S.~An, L.~P.~Zhang and H.~S.~Zong,
  %``Finite volume effects on the chiral phase transition from Dyson–Schwinger equations of QCD,''
  Nucl.\ Phys.\ B {\bf 938}, 298 (2019).
%  doi:10.1016/j.nuclphysb.2018.11.015
%  [arXiv:1711.04914 [hep-ph]].
  %%CITATION = doi:10.1016/j.nuclphysb.2018.11.015;%%
  %5 citations counted in INSPIRE as of 29 Dec 2018

%\cite{Shi:2018swj}
\bibitem{Shi:2018swj} 
  C.~Shi, W.~Jia, A.~Sun, L.~Zhang and H.~Zong,
  %``Chiral crossover transition in a finite volume,''
  Chin.\ Phys.\ C {\bf 42}, no. 2, 023101 (2018).
  doi:10.1088/1674-1137/42/2/023101
  %%CITATION = doi:10.1088/1674-1137/42/2/023101;%%
  %2 citations counted in INSPIRE as of 29 Dec 2018
 
 %\cite{Braun:2004yk}
\bibitem{Braun:2004yk} 
  J.~Braun, B.~Klein and H.-J.~Pirner,
  %``Volume dependence of the pion mass in the quark-meson-model,''
  Phys.\ Rev.\ D {\bf 71}, 014032 (2005)
  doi:10.1103/PhysRevD.71.014032
  [hep-ph/0408116].
  %%CITATION = doi:10.1103/PhysRevD.71.014032;%%
  %38 citations counted in INSPIRE as of 29 Dec 2018
  
  %\cite{Braun:2005fj}
\bibitem{Braun:2005fj} 
  J.~Braun, B.~Klein, H.-J.~Pirner and A.~H.~Rezaeian,
  %``Volume and quark mass dependence of the chiral phase transition,''
  Phys.\ Rev.\ D {\bf 73}, 074010 (2006)
  doi:10.1103/PhysRevD.73.074010
  [hep-ph/0512274].
  %%CITATION = doi:10.1103/PhysRevD.73.074010;%%
  %43 citations counted in INSPIRE as of 29 Dec 2018
 
 %\cite{Ferrer:1999gs}
\bibitem{Ferrer:1999gs} 
  E.~J.~Ferrer, V.~P.~Gusynin and V.~de la Incera,
  %``Boundary effects in the magnetic catalysis of chiral symmetry breaking,''
  Phys.\ Lett.\ B {\bf 455}, 217 (1999)
  doi:10.1016/S0370-2693(99)00470-0
  [hep-ph/9901446].
  %%CITATION = doi:10.1016/S0370-2693(99)00470-0;%%
  %26 citations counted in INSPIRE as of 26 Feb 2019
  
 \bibitem{Abreu:2006}
  L.~M.~Abreu, M.~Gomes and A.~J.~da~Silva,
  %``Finite-size effects on the phase diagram of difermion condensates in two-dimensional four-fermion interaction models,''
  {\it  Phys. Lett. B } {\bf 642}, 551 (2006).

\bibitem{Ebert0} D. Ebert, K. G. Klimenko, A. V. Tyukov and V. Ch. Zhukovsky, {\it Phys. Rev. D} {\bf 78}, 045008 (2008).

%\cite{Abreu:2009zz}
\bibitem{Abreu:2009zz}
  L.~M.~Abreu, A.~P.~C.~Malbouisson, J.~M.~C.~Malbouisson and A.~E.~Santana,
  %``Finite-size effects on the chiral phase diagram of four-fermion models in four dimensions,''
   {\it Nucl.\ Phys.\ B } {\bf 819}, 127 (2009).
%  [arXiv:0909.5105 [hep-th]].
  %%CITATION = ARXIV:0909.5105;%%
  
   %\cite{Abreu:2011rj}
\bibitem{Abreu:2011rj}
  L.~M.~Abreu, A.~P.~C.~Malbouisson and J.~M.~C.~Malbouisson,
  %``Finite-size effects on the phase diagram of difermion condensates in two-dimensional four-fermion interaction models,''
  {\it  Phys.\ Rev.\ D } {\bf 83}, 025001 (2011).
 % [arXiv:1102.1860 [hep-th]].
  %%CITATION = ARXIV:1102.1860;%%

 %\cite{Bhattacharyya:2012rp}
\bibitem{Bhattacharyya:2012rp} 
  A.~Bhattacharyya, P.~Deb, S.~K.~Ghosh, R.~Ray and S.~Sur,
  %``Thermodynamic Properties of Strongly Interacting Matter in Finite Volume using Polyakov-Nambu-Jona-Lasinio Model,''
  Phys.\ Rev.\ D {\bf 87}, no. 5, 054009 (2013).
%  doi:10.1103/PhysRevD.87.054009
%  [arXiv:1212.5893 [hep-ph]].
  %%CITATION = doi:10.1103/PhysRevD.87.054009;%%
  %36 citations counted in INSPIRE as of 29 Dec 2018
  
%\cite{Bhattacharyya:2014uxa}
\bibitem{Bhattacharyya:2014uxa} 
  A.~Bhattacharyya, R.~Ray and S.~Sur,
  %``Fluctuation of strongly interacting matter in the Polyakov–Nambu–Jona-Lasinio model in a finite volume,''
  Phys.\ Rev.\ D {\bf 91}, no. 5, 051501 (2015).
%  doi:10.1103/PhysRevD.91.051501
%  [arXiv:1412.8316 [hep-ph]].
  %%CITATION = doi:10.1103/PhysRevD.91.051501;%%
  %25 citations counted in INSPIRE as of 29 Dec 2018

\bibitem{Bhattacharyya2} 
 A. Bhattacharyya, R. Ray, S. Samanta and S. Sur, {\it Phys. Rev. C} {\bf 91}, 041901 (2015).

%\cite{Kohyama:2016fif}
\bibitem{Kohyama:2016fif} 
  H.~Kohyama, D.~Kimura and T.~Inagaki,
  %``Parameter fitting in three-flavor Nambu–Jona-Lasinio model with various regularizations,''
  Nucl.\ Phys.\ B {\bf 906}, 524 (2016)
%  doi:10.1016/j.nuclphysb.2016.03.015
%  [arXiv:1601.02411 [hep-ph]].
  %%CITATION = doi:10.1016/j.nuclphysb.2016.03.015;%%
  %2 citations counted in INSPIRE as of 22 Feb 2019
  
%\cite{Pan:2016ecs}
\bibitem{Pan:2016ecs} 
  Z.~Pan, Z.~F.~Cui, C.~H.~Chang and H.~S.~Zong,
  %``Finite-volume effects on phase transition in the Polyakov-loop extended Nambu–Jona-Lasinio model with a chiral chemical potential,''
  Int.\ J.\ Mod.\ Phys.\ A {\bf 32}, no. 13, 1750067 (2017).
%  doi:10.1142/S0217751X17500671
%  [arXiv:1611.07370 [hep-ph]].
  %%CITATION = doi:10.1142/S0217751X17500671;%%
  %9 citations counted in INSPIRE as of 29 Dec 2018
  
\bibitem{Wang:2018qyq} 
  Q.~Wang, Y.~Xiq and H.~Zong,
  %``Nambu–Jona-Lasinio model with proper time regularization in a finite volume,''
  Mod.\ Phys.\ Lett.\ A {\bf 33}, no. 39, 1850232 (2018)
%  doi:10.1142/S0217732318502322
%  [arXiv:1806.05315 [hep-ph]].
  %%CITATION = doi:10.1142/S0217732318502322;%%
  %1 citations counted in INSPIRE as of 26 Feb 2019  
     
   
%\cite{Gasser:1986vb}
\bibitem{Gasser:1986vb} 
  J.~Gasser and H.~Leutwyler,
  %``Light Quarks at Low Temperatures,''
  Phys.\ Lett.\ B {\bf 184}, 83 (1987).
%  doi:10.1016/0370-2693(87)90492-8
  %%CITATION = doi:10.1016/0370-2693(87)90492-8;%%
  %642 citations counted in INSPIRE as of 29 Dec 2018


 %\cite{Damgaard:2008zs}
\bibitem{Damgaard:2008zs} 
  P.~H.~Damgaard and H.~Fukaya,
  %``The Chiral Condensate in a Finite Volume,''
  JHEP {\bf 0901}, 052 (2009)
%  doi:10.1088/1126-6708/2009/01/052
%  [arXiv:0812.2797 [hep-lat]].
  %%CITATION = doi:10.1088/1126-6708/2009/01/052;%%
  %42 citations counted in INSPIRE as of 29 Dec 2018


 \bibitem{Fraga}  E. S. Fraga and L. F.  Palhares, Phys. Rev. D {\bf 86}, 016008 (2012). 


\bibitem{Abreu3} L. M. Abreu, C. A. Linhares, A. P. C. Malbouisson, J. M. C. Malbouisson,  {\it Phys. Rev. D} {\bf  88}, 107701 (2013).

\bibitem{Abreu6}L. M. Abreu, A. P. C. Malbouisson, J. M. C. Malbouisson, E. S. Nery and R. Rodrigues da Silva, Nucl. Phys. B \textbf{881}, 327-342 (2014).

\bibitem{Ebert3} D. Ebert, T. G. Khunjua, K. G. Klimenko and V. C. Zhukovsky,  {\it Phys. Rev. D} {\bf 91}, 105024 (2015). 


 \bibitem{Abreu4} L. M. Abreu, E. S. Nery and A. P. C. Malbouisson, Phys. Rev. D \textbf{91}, 087701 (2015).

\bibitem{Abreu5} L. M. Abreu and E. S. Nery, Int. J. Mod. Phys. A {\bf 31}, 1650128 (2016). 

\bibitem{Abreu7} L. M. Abreu, A. P. C. Malbouisson and E. S. Nery, Mod. Phys. Lett. A {\bf 31}, 1650121 (2016).

\bibitem{Bao1} S.S. Bao and H. Shen, Phys. Rev. C {\bf 93}, 025807 (2016).
%40
\bibitem{PhysRevC.96.055204} L. M. Abreu, and E. S. Nery, {\it  Phys.\ Rev.\ C }
 {\bf 96}, 055204 (2017).  

\bibitem{Samanta} S. Samanta, S. Ghosh and B. Mohanty, J. Phys. G: Nucl. Part. Phys. {\bf 45}, 075101 (2018).

\bibitem{Wu} X.H. Wu and H. Shen, Phys. Rev. C \textbf{96}, 025802 (2017).

%\cite{Klein:2017shl}
\bibitem{Klein:2017shl} 
  B.~Klein,
  %``Modeling Finite-Volume Effects and Chiral Symmetry Breaking in Two-Flavor QCD Thermodynamics,''
  Phys.\ Rept.\  {\bf 707-708}, 1 (2017).
%  doi:10.1016/j.physrep.2017.09.002
%  [arXiv:1710.05357 [hep-ph]].
  %%CITATION = doi:10.1016/j.physrep.2017.09.002;%%
  %6 citations counted in INSPIRE as of 29 Dec 2018


\bibitem{Shi} C. Shi, Y. Xia, W. Jia et al., Sci. China Phys. Mech. Astron.  {\bf 61} 082021 (2018). 


\bibitem{Kharzeev} D.E Kharzeev, L.D. Mclerran and H.J. Warringa, ArXiv:0711.0950 [hep-ph], 2007.
  
\bibitem{Skokov:2009qp} V.~Skokov, A.~Y.~Illarionov and V.~Toneev, Int.\ J.\ Mod.\ Phys.\ A {\bf 24}, 5925 (2009).  

 %\cite{Chernodub:2010qx}
\bibitem{Chernodub:2010qx} M.~N.~Chernodub,
  %``Superconductivity of QCD vacuum in strong magnetic field,''
  Phys.\ Rev.\ D {\bf 82}, 085011 (2010).
 % doi:10.1103/PhysRevD.82.085011
 % [arXiv:1008.1055 [hep-ph]].
  %%CITATION = doi:10.1103/PhysRevD.82.085011;%%
  %179 citations counted in INSPIRE as of 15 Dec 2018


\bibitem{Ayala1} A. Ayala, M. Loewe, J.C. Rojas and C. Villavicencio, Phys. Rev. D \textbf{86}, 076006 (2012).


 \bibitem{Tobias} M. Ferreira, P. Costa, O. Lourenco, T. Frederico, and C. Providencia,  Phys. Rev. D {\bf 89}, 116011 (2014).
  
\bibitem{Heber} A. Haber, F. Preis and A. Schmitt, Phys. Rev. D \textbf{90}, 125036 (2014).

\bibitem{MAO} S. Mao, Phys. Lett. B {\bf 758}, 195 (2016).

%50
\bibitem{Ayala2} A. Ayala, P. Mercado and C. Villavicencio, Phys. Rev. C \textbf{95}, 014904 (2017).





\bibitem{Mamo:2015dea}  K.~A.~Mamo,
  %``Inverse magnetic catalysis in holographic models of QCD,''
  JHEP {\bf 1505}, 121 (2015)
%  doi:10.1007/JHEP05(2015)121
%  [arXiv:1501.03262 [hep-th]].
  %%CITATION = doi:10.1007/JHEP05(2015)121;%%
  %40 citations counted in INSPIRE as of 14 Dec 2018

\bibitem{Pagura} V. P. Pagura, D. Gomez Dumm, S. Noguera, and N. N. Scoccola
Phys. Rev. D {\bf 95}, 034013 (2017).

\bibitem{Magdy} N. Magdy, M. Csanad and R. A. Lacey, J. Phys. G: Nucl. Part. Phys. {\bf 44}, 025101 (2017).

\bibitem{Ayala0} A. Ayala, C. A. Dominguez, S. Hernandez-Ortiz, L. A. Hernandez, M. Loewe, D. M. Paret, and R. Zamora, Phys. Rev. D {\bf 98}, 031501(R) (2018).
   
  
\bibitem{NJL} Y. Nambu and G. Jona-Lasinio, Phys. Rev. \textbf{122}, 345 (1961).

\bibitem{NJL1} Y. Nambu and G. Jona-Lasinio, Phys. Rev. \textbf{124}, 246 (1961).

\bibitem{Vogl} U. Vogl and W. Weise, Prog. Part. Nucl. Phys. \textbf{27}, 195 (1991).
	
\bibitem{Klevansky} S. P. Klevansky, Rev. Mod. Phys. {\bf 27}, 195 (1991).

\bibitem{Hatsuda} T. Hatsuda and T. Kunikiro, Phys. Rep. \textbf{247}, 221 (1994).

\bibitem{Buballa} M. Buballa, Phys. Rep. \textbf{407}, 205 (2005).

%\bibitem{Luciano} L.M. Abreu, A.P.C. Malbouisson, J.M.C. Malbouisson, and A.E. Santana, arXiv:0909.5105v1 [hep-th] (2009).

\bibitem{livro} F.C. Khanna, A.P.C. Malbouisson, J.M.C. Malbouisson, and A.E.
Santana, {\it{Thermal Quantum Field Theory: Algebraic Aspects and Applications}}%
, World Scientific, Singapore (2009).

\bibitem{PR2014} F.C. Khanna, A.P.C. Malbouisson, J.M.C. Malbouisson, and A.E.
Santana,, Phys. Rep. \textbf{539}, 135 (2014). 

\bibitem{Emerson} E. B. S. Corr\^ea, C. A. Linhares, A. P. C. Malbouisson, J. M. C. Malbouisson, and A. E. Santana, Eur. Phys. J. C, \textbf{77}, 261 (2017).
	
\bibitem{Bellac} M. Le Bellac, \textit{Thermal Field Theory} (Cambridge University 
Press, Cambridge, UK, 1996).

\bibitem{Kapusta} J.I. Kapusta and C. Gale, \textit{Finite-Temperature Field Theory: Principles
and Applications} (Cambridge University Press, Cambridge, UK, 2006). 


%\bibitem{Chineses} Qing-Wu Wang, Younghui Xia, and Hong-Shi Zong, arXiv:1806.05315v1 [hep-ph] 14 Jun (2018).

\bibitem{Schwinger} J. Schwinger, Phys. Rev. \textbf{82}, 914 (1951).

\bibitem{Farina} F.A. Barone, H. Boschi-Filho, C. Farina, Am. J. Phys. \textbf{71}, 483 (2003).


\bibitem{Bellman} R. Bellman, {\it{A Brief Introduction to Theta Functions}}, Holt, Rinehart and Winston, Inc., New York (1961).

%\bibitem{Ritus} V. I. Ritus, Ann. Phys. \textbf{69}, 555 (1972).
%
%\bibitem{Emerson1} E. B. S. Corrêa and J. E. Oliveira, Rev. Bras. Ens. Fís, \textbf{37}, 3302 (2015).


\bibitem{Franceses} 
%F. Gastineau, R. Nebauer, and J. Aichelin, arXiv:hep-ph/0101289v3, 20 Nov (2001).
%
%\cite{Nebauer:2001rb}
%\bibitem{Nebauer:2001rb} 
  F.~Gastineau, R.~Nebauer and J.~Aichelin,
  %``Thermodynamics of the three flavor NJL model: Chiral symmetry breaking and color superconductivity,''
  Phys.\ Rev.\ C {\bf 65}, 045204 (2002).
%  doi:10.1103/PhysRevC.65.045204
%  [hep-ph/0101289].
  %%CITATION = doi:10.1103/PhysRevC.65.045204;%%
  %55 citations counted in INSPIRE as of 14 Jan 2019

%\bibitem{ProvidenciaMagnetico} D.P. Menezes, M. Benghi Pinto, S.S. Avancini, and C. Provid\^encia, arXiv:0907.2607v1 [nucl-th] 15 Jul (2009).
%
%

%\bibitem{Taiwan} H. Kohyama, arXiv:1602.09056v1 [hep-ph] (2016).



%\bibitem{Birrel} N.D. Birrell and L.H. Ford, Phys. Rev. D \textbf{22}, 330 (1980).

%\bibitem{AOP11} F.C. Khanna, A.P.C. Malbouisson, J.M.C. Malbouisson, and A.E. Santana, Ann. Phys. \textbf{326}, 2364 (2011).

%%%%%%%%%%%%%%%%%%%%%%%%%%%%%%%%%%%%%%%%%%%%%%%%%%%%%%%%%%%%%%%%%%%%%%%%%%%%%%%%%%%%%%%%%%%%%%%%%%%%%%%%%%%%%%%%%%%%%%%%%%%%%%%%%%%%%%%%%%%%%%%%%%%%%%%%%%%%%%%%%%%%%%%%%%%%%%%%%%%%%%%%%%%%%%%%%%%%%%%%%%%%%%%%%%%%%%%%%%%%%%%%%%%%%%%%%%%%%%%%%%%%%%%%%%%%%%%%%%%%%%%%%%%%%%%%%%%%%%%%%%

\end{thebibliography}
\end{document}